\documentclass[10pt,aps,prd,nofootinbib,superscriptaddress,twocolumn]{revtex4}
\usepackage[utf8]{inputenc}
\DeclareUnicodeCharacter{200B}{{\hskip 0pt}}
\usepackage{color}
\usepackage{graphicx}
\usepackage{amsmath}\allowdisplaybreaks
\usepackage{amssymb}
\usepackage{bm}
\usepackage{hyperref}
\usepackage{comment}
\hypersetup{
    colorlinks=true,
    linkcolor=blue,
    filecolor=magenta,      
    urlcolor=black,
		citecolor=red,
		}
\begin{document}

\title{Maxwell equations in Schwarzschild spacetime for static and freely falling observers}

\author{F. L. Carneiro}\email{fernandolessa45@gmail.com}
\affiliation{Universidade Federal do Norte do Tocantins, 77824-838, Aragua\'ina, TO, Brazil}

\author{L. V. A. Cunha}\email{l.v.a.cunha100@gmail.com}
\affiliation{Universidade Federal do Norte do Tocantins, 77824-838, Aragua\'ina, TO, Brazil}

\date{\today}

\begin{abstract}
Classical electrodynamics in curved spacetime is formulated within a tetrad-based framework that preserves the direct physical interpretation of the electromagnetic fields measured by an observer. The formalism is applied to Schwarzschild spacetime for two distinct families of observers described in the same coordinate system: static observers and radially free-falling observers. For static observers, Maxwell equations retain their usual spherical structure, while the gravitational field introduces geometrical corrections in the radial and temporal sectors through the metric function. These corrections are interpreted in terms of proper radial distance, proper time, and the radial variation of the lapse function. For radially free-falling observers, additional kinematical contributions arise from the local radial boost relating the free-falling frame to the static one. As a consequence, charge density mixes with radial current, and the electric and magnetic sectors become intertwined in the temporal-radial projections of Maxwell equations, whereas the angular Amp\`ere-Maxwell and Faraday equations retain the same structure found for the static observer. The weak-field and near-horizon regimes are also examined, and the results are discussed through an effective-medium analogy, in which the Schwarzschild geometry behaves as an inhomogeneous geometrical medium at rest for static observers and as a radially moving effective medium in the temporal-radial sector for free-falling observers.
\end{abstract}


\maketitle
\section{Introduction}

The unification of electric and magnetic phenomena by Maxwell at the end of the 19th century, as part of the same electromagnetic phenomenon, prompted the mathematical characterization of the electric $\mathbf{E}$ and magnetic $\mathbf{B}$ fields as components of a rank-two tensor, named the Faraday tensor $F^{\mu\nu}$ \cite{minkowski1908grundgleichungen, hehl2008maxwell}.

The apparent necessity of a privileged inertial frame, in the perception of Lorentz and others, for the validity of the evolution equations of the theory, i.e., Maxwell's equations, culminated in extensive experimental investigations on the existence of such a frame (a medium called the \textit{aether luminiferus}). However, the failure of the expected behavior, as demonstrated by the Michelson--Morley and Fizeau experiments, culminated in the \textit{ad hoc} definition of the Lorentz transformations. Such transformations were introduced to comply with electromagnetic theory, in contrast with the Galilean transformations.

Soon after, Einstein proposed the theory of Special Relativity (SR) in 1905. With this new framework, the existence of such a medium became irrelevant. The fundamental axioms of SR already incorporate the Lorentz transformations \textit{ab initio}. Hence, Maxwell's theory was already relativistic even before the formal development of SR.

However, although Maxwell electrodynamics (ME) is a relativistic theory, and Relativistic Mechanics was later developed encompassing Newtonian mechanics as a limiting case, Newtonian gravity remains essentially non-relativistic.

In order to resolve the incompatibility of the gravitational interaction with his new theory, Einstein abandoned the notion of force and modeled gravitation as a manifestation of spacetime geometry, a concept that, although optional in SR, is essential in General Relativity (GR). In this framework, spacetime is described as a four-dimensional manifold endowed with a symmetric (torsion-free) and metric-compatible connection, whose nontrivial structure is encoded in the Riemann curvature tensor.

The emergence of GR raised another issue, namely that ME had been formulated in flat Minkowski spacetime. The departure from flatness in the presence of gravity requires adjustments to the theory, from the evolution equations to the interpretation of the components of $F^{\mu\nu}$ as the physically measured electric $\tilde{E}_{i}$ and magnetic $\tilde{B}_{i}$ fields \cite{hwang2023definition}. While such an interpretation is straightforward in flat spacetime, the presence of nontrivial geometry makes it more difficult to assign direct physical meaning to these quantities.

Einstein initially pursued a unification of electromagnetic and gravitational phenomena but was unsuccessful in deriving both Maxwell's and Einstein equations from a single Lagrangian. To this day, no approach to this problem stands out. Consequently, the literature typically treats the electromagnetic field as a physical field defined over a curved spacetime, i.e., through a coupling between the electromagnetic field and spacetime geometry \cite{aly2025coupling}. Several levels of coupling may be considered, the most fundamental being the replacement of ordinary derivatives by covariant ones in Maxwell's equations, without explicit dependence on the curvature tensor.

When coupling the electromagnetic field, a fundamental question arises: to which geometry should it be coupled? It is common in the literature to adopt Riemannian geometry, as in Refs.~\cite{frolov2020maxwell,mavrogiannis2021electromagnetic,formiga2023gravitational}. While this is a valid approach, we show in this article that there are advantages in coupling the electromagnetic field to Weitzenb\"{o}ck geometry \cite{weitzenbock1932invarianten}. First, this framework provides a consistent coupling between the gravitational and electromagnetic fields, since it is possible to construct a gravitational theory --- the Teleparallel Equivalent of General Relativity (TEGR) \cite{maluf2013teleparallel} --- in which the spacetime dynamics are equivalent to those of GR; hence, all known solutions of GR are also solutions of TEGR. Second, the emergence of tetrad fields allows a clear separation between coordinate and frame indices, providing an elegant relation between the physical electromagnetic fields and the components of the Faraday tensor. The independence between coordinate and frame transformations enables a generalization of Maxwell's equations to arbitrary coordinate systems and observers.

In this geometry, spacetime is endowed with torsion associated with the affine connection and curvature associated with the Levi-Civita connection, both related through the contortion tensor. This feature is particularly useful when describing electrodynamics in accelerated frames \cite{maluf2010electrodynamics}, since gravity gives rise to both torsion (of the affine connection) and curvature (of the Levi-Civita connection), whereas acceleration alone of the observer gives rise only to torsion. Therefore, electrodynamics may be consistently generalized to non-inertial frames. Interestingly, the aforementioned reference shows that acceleration possesses an absolute character from a physical standpoint: a static observer measures nonzero radiation from an accelerated charge, while an accelerated observer measures no radiation from a static charge.

The role of the observer in the measurement of electromagnetic fields becomes particularly subtle when noninertial effects are taken into account. In fact, realistic observers are necessarily accelerated, and the standard formulation of electrodynamics relies on the hypothesis of locality, according to which an accelerated observer is instantaneously equivalent to a comoving inertial one. However, this assumption is only approximate and may break down for wave phenomena, where intrinsic nonlocal effects become relevant \cite{bremm2015nonlocal}. This limitation has motivated the development of nonlocal extensions of electrodynamics, in which the measured fields depend on the past history of the observer's motion. From a complementary perspective, the generally covariant formulation of Maxwell's theory shows that no modification of the field equations is required in noninertial frames; instead, all inertial and gravitational effects are encoded in the choice of frame and in the associated tetrad fields that define the physical electric and magnetic components \cite{obukhov2021electrodynamics}. In this context, the construction of proper reference frames for accelerated observers, together with the associated tetrad formalism, is important in consistently defining measurements and interpreting physical quantities \cite{formiga2014accelerated}. 

In this paper, we consider the coupling between the electromagnetic field and spacetime described by Weitzenb\"{o}ck geometry, in a formulation valid for arbitrary observers in arbitrary spacetimes, and which allows a clear interpretation of the physical electromagnetic fields. This is achieved by projecting the components of the Faraday tensor onto the observer's frame, geometrically characterized by the tangent Minkowski spacetime at the measurement event. As a first application, we restrict our analysis to Schwarzschild spacetime, although the procedure remains valid in general. Furthermore, taking advantage of the independence between coordinate and frame transformations, we analyze two distinct reference frames. First, we consider a static observer in Schwarzschild coordinates and compute the four vectorial laws of electromagnetism. Second, we consider a radially free-falling observer in the same coordinate system and perform the same analysis. By comparing both cases, we observe that the radially dependent components of Maxwell's equations are strongly influenced by the absence of inertial acceleration in the frame. Additionally, we show that the gravitational field behaves as an effective medium for the electromagnetic field, modifying the effective electric permittivity in Gauss's law.

Our article is organized as follows. In Sec.~\ref{sec2}, we review the main aspects of Weitzenb\"{o}ck geometry and introduce the Mashhoon acceleration tensor, which is used to characterize the inertial properties of the chosen reference frame. In Sec.~\ref{sec3}, we present the construction of the Faraday tensor in curved spacetime in terms of locally measured electric and magnetic fields. We also discuss how the preservation of the physical interpretation of its components leads naturally to a connection compatible with the parallel transport of local quantities. In Sec.~\ref{sec4}, we evaluate Maxwell's equations in Schwarzschild spacetime for a static observer. In this case, the equations retain the usual spherical structure, while the gravitational field modifies the radial and temporal sectors through the metric factors associated with proper distance and proper time. In Sec.~\ref{sec5}, we repeat the analysis for a radially free-falling observer. We show that this frame is related to the static one by a local radial boost, which mixes charge density with radial current and mixes the electric and magnetic sectors in the temporal-radial projections of Maxwell equations, whereas the angular Amp\`ere-Maxwell and Faraday equations retain the same structure found for the static observer. In Sec.~\ref{sec6}, we analyze these results from the point of view of observer dependence, separating the geometrical corrections of the static frame from the kinematical effects produced by the free-fall boost. We show that the gravitational field admits an effective medium interpretation for the electromagnetic field, modifying the radial electromagnetic response encoded in Gauss's law. Finally, in Sec.~\ref{conc}, we present our conclusions.

Notation: spacetime indices $\mu,\nu,\ldots$ take values $0,1,2,3$, while Lorentz indices $a,b,\ldots$ take values $(0),(1),(2),(3)$. The spacetime metric $g_{\mu\nu}$ is used to raise and lower spacetime indices, and the Minkowski metric $\eta_{ab}$ is used to raise and lower Lorentz indices. We adopt the signature $(-,+,+,+)$ and geometrized units, with electromagnetic constants absorbed into the definition of the current.

\section{Observers in spacetime}\label{sec2}

We may interpret spacetime as a metric 4D manifold $\mathcal{M}$ endowed, at first, with properties such as curvature, torsion and non-metricity, described by coordinates $x^{\mu}$ that label the points of that manifold. Tangent to each event $P$, we may associate a flat tangent spacetime described by the Latin indices at the beginning of the alphabet, e.g., $a,b,c,\ldots$. In this tangent space $T_{P}\mathcal{M}$ we therefore have a set of four linearly independent vectors $e_{a}{}^{\mu}=\{e_{(0)}{}^{\mu},e_{(i)}{}^{\mu}\}$, where the spacetime index denotes the event. When we consider the whole spacetime, we have a vector field.

The vector field components $e_{a}{}^{\mu}$ are known as tetrad fields. The tetrads allow the projection of components of $T_{P}\mathcal{M}$ into spacetime, e.g., for a vector $v^{a}$ we have $v^{\mu}=e_{a}{}^{\mu}v^{a}$. Analogously, for the cotangent space we have $e^{a}{}_{\mu}$ and we may project the other way around, e.g., $v^{a}=e^{a}{}_{\mu}v^{\mu}$. Hence, the orthogonality condition for the tetrads implies
\begin{equation}\label{flatmetric}
	e_{a}{}^{\mu}e_{b\mu}=\eta_{ab}\,,
\end{equation}
with $\eta_{ab}=\mathrm{diag}(-1,1,1,1)$ denoting the metric tensor of the flat tangent space. Also, we may obtain the metric tensor of spacetime from the tetrads as
\begin{equation}\label{spacetimemetric}
	g_{\mu\nu}=e^{a}{}_{\mu}e_{a\nu}\,.
\end{equation}

Since the tangent space is always flat and independent of the coordinates describing spacetime, projected quantities into it have the same physical meaning as in Minkowski spacetime and are invariant under coordinate transformations, although still observer dependent.

Let us consider an observer in spacetime with worldline $x^{\mu}(\tau)$, where $\tau$ is its proper time. At each event $x^{\mu}$, the observer has an Instantaneous Rest Frame (IRF) where, infinitesimally, it is at rest, i.e., a frame that can be identified with the tangent space at $x^{\mu}$. Hence, when measuring some quantity, the observer projects it into its IRF, i.e., into the tangent space at that event. Thus, we may treat the vector $v^{a}$ as having direct physical meaning. For example, when considering the four-momentum of a particle $p^{\mu}$, the observer measures the coordinate-independent energy $p^{(0)}=e^{(0)}{}_{\mu}\,p^{\mu}$. The measurement is, however, observer dependent, but this is not a problem, since the observer has a physical nature and it is expected that physical quantities depend on the observer measuring them.

The characterization of the observer can be constructed by initially considering the observer's four-velocity $u^{\mu}$. Given that $e_{a}{}^{\mu}=\{e_{(0)}{}^{\mu},e_{(i)}{}^{\mu}\}$ represents the IRF, we know that the observer is always at rest, thus $e_{(0)}{}^{\mu}$ is tangent to its worldline, making it possible to identify
\begin{equation}\label{relationvelandtetrads}
	u^{\mu}=e_{(0)}{}^{\mu}\,.
\end{equation}
Hence, from \eqref{relationvelandtetrads}, the four-acceleration of the observer is given by
\begin{equation}\label{4acce}
    a^\mu(\tau)=\dfrac{D e_{(0)}{}^\mu}{d\tau}=u^\alpha\mathring{\nabla}_\alpha e_{(0)}{}^\mu\,,
\end{equation}
where the circle indicates a covariant derivative with respect to the Christoffel symbols $\mathring\Gamma^\mu{}_{\alpha\beta}$.

Until now we have a general result valid for any chosen geometry. Let us now restrict ourselves further. Given the ability of the tetrads to carry information about spacetime from \eqref{spacetimemetric} and about the observer from \eqref{4acce}, it is of interest to consider a tetrad-based geometric formulation. In addition, the necessity to characterize accelerated frames makes it useful to consider a geometry endowed with torsion, since the torsion may be used to encode both gravitational effects and inertial properties of the frame in spacetime. One of such geometries is the one considered by Weitzenb\"{o}ck in Ref.~\cite{weitzenbock1932invarianten}. Although the nomenclature and development differ from the modern approach, an interesting consideration was developed: a geometry that guarantees the parallel transport of vectors at a distance independently of the path, i.e., a teleparallel geometry with vanishing curvature for the affine connection (and we shall add with metricity). Hence, in such geometry, the tetrads are parallel transported, thus the affine connection of the theory $\Gamma^{\lambda}{}_{\mu\nu}$ yields a vanishing covariant derivative for the tetrads:
\begin{equation}\label{parallel}
	\nabla_{\mu}e^{a}{}_{\nu}=\partial_{\mu}e^{a}{}_{\nu}-\Gamma^{\lambda}{}_{\mu\nu}e^{a}{}_{\lambda}=0\,.
\end{equation}
From the above equation, we may write the connection as
\begin{equation}\label{weitzenbock}
	\Gamma^{\lambda}{}_{\mu\nu}=e_{a}{}^{\lambda}\partial_{\mu}e^{a}{}_{\nu}\,,
\end{equation}
yielding a non-vanishing torsion tensor
\begin{equation}
	T^{a}{}_{\mu\nu}=\partial_{\mu}e^{a}{}_{\nu}-\partial_{\nu}e^{a}{}_{\mu}\,.
\end{equation}
From the torsion tensor we may compute the contorsion tensor
\begin{equation}\label{contorsion}
	K_{\mu ab} = \frac{1}{2}\, e_{a}{}^{\lambda}\, e_{b}{}^{\nu} \left( T_{\lambda\mu\nu} + T_{\nu\lambda\mu} + T_{\mu\lambda\nu} \right)\,.
\end{equation}
Since the contorsion is the difference between two connections, we have
\begin{equation}
	K_{\mu ab} = \omega_{\mu ab}-\mathring{\omega}_{\mu ab}\,,
\end{equation}
where $\omega_{\mu ab}$ is the spin connection and $\mathring{\omega}_{\mu ab}$ is the torsion-free Levi-Civita connection.

Although not the objective of the present paper, by constructing the appropriate Weitzenb\"{o}ck invariants, a theory dynamically invariant to GR, namely TEGR, can be constructed in Weitzenb\"{o}ck geometry, where we have the same gravitational solutions, since TEGR and GR field equations are equivalent. It can be seen in Eq.~(16) of Ref.~\cite{maluf2005gravitational} that the connection $\omega_{\mu ab}$ plays no role in the dynamics of the field. Hence, we may consider it as zero and write the Levi-Civita connection from the torsion tensor as
\begin{equation}\label{levi}
	-\mathring{\omega}_{\mu ab}=K_{\mu ab}=\frac{1}{2}\, e_{a}{}^{\lambda}\, e_{b}{}^{\nu} \left( T_{\lambda\mu\nu} + T_{\nu\lambda\mu} + T_{\mu\lambda\nu} \right)\,.
\end{equation}
We could treat the connection as non-vanishing and use it as a gauge structure, but such an approach leads to several problems \cite{maluf2020difficulties}.

Let us now return to the observer's characterization problem by considering the spacetime as a Weitzenb\"{o}ck manifold. Let us consider the total derivative of the tetrads with respect to the observer's proper time projected into the tangent space
\begin{equation}\label{phi}
	\phi_{a}{}^{b} := e^{b}{}_{\mu}\frac{D e_{a}{}^{\mu}}{d\tau}=e^{b}{}_{\mu}u^{\nu} \partial_{\nu} e_{a}{}^{\mu}+ e^{b}{}_{\mu}u^{\nu}\mathring{\Gamma}^\mu{}_{\nu\lambda} e_{a}{}^{\lambda}\,.
\end{equation}
Since we may also compute the covariant derivative of the tetrads from the Christoffel symbols and the Levi-Civita connection, we have \cite{maluf2013teleparallel}
\begin{equation}\label{relation}
	\partial_{\nu}e^{b}{}_{\lambda}-\mathring{\Gamma}^{\sigma}{}_{\nu\lambda}e^{b}{}_{\sigma}+\mathring{\omega}_{\nu}{}^{b}{}_{c}e^{c}{}_{\lambda}=0\,.
\end{equation}
By noticing that the first term in the RHS of \eqref{phi} can be written using \eqref{relation}, we may write \eqref{phi} as
\begin{equation}\label{mashhoon}
	\phi_{a}{}^{b}=e_{(0)}{}^{\lambda}\,\mathring{\omega}_{\lambda}{}^{b}{}_{a}=\frac{1}{2} \left[ T_{(0)ab} + T_{a(0)b} - T_{b(0)a} \right]\,.
\end{equation}
As $a^{(i)}=e^{(i)}{}_{\mu}a^{\mu}=\phi_{(0)}{}^{(i)}$, we may identify the components $\phi_{(0)}{}^{(i)}$ as the translational acceleration of the frame in the $(i)$ direction and $\phi_{(i)}{}^{(j)}$ as the frame rotation with respect to a Fermi-Walker transported frame \cite{mashhoon2003vacuum}.

The antisymmetric acceleration tensor $\phi_{a}{}^{b}$ is known as the Mashhoon tensor and allows the characterization of the reference frame in a given spacetime. It can be interpreted as the inertial acceleration required to maintain the frame in its kinematical state, with four-velocity given by $u^{\mu}=e_{(0)}{}^{\mu}$. Therefore, the torsion tensor has a significant importance in the dynamics of the observer.

We shall use the Mashhoon tensor to interpret the tetrads employed in the computation of Maxwell's equations.	

\section{Maxwell equations in Schwarzschild spacetime}\label{sec3}

The construction of Maxwell equations in an arbitrary coordinate system requires first the construction and interpretation of the Faraday tensor in such coordinates. One could work with the four-potential, but here we aim to keep the physical interpretation of the fields, since we are interested in the measurement of electric and magnetic fields by real observers. Hence, we depart from the Faraday tensor in the local Lorentz basis $F^{ab}$, i.e., projected into the frame of reference of some observer, in order to assign a direct physical interpretation to its components. In Cartesian-like coordinates this tensor reads
\begin{equation}\label{faradaylocal}
F_{ab}=
\begin{pmatrix}
0 & \tilde{E}_x & \tilde{E}_y & \tilde{E}_z \\\\
-\tilde{E}_x & 0 & -\tilde{B}_z & \tilde{B}_y \\\\
-\tilde{E}_y & \tilde{B}_z & 0 & -\tilde{B}_x \\\\
-\tilde{E}_z & -\tilde{B}_y& \tilde{B}_x &0
\end{pmatrix}\,,
\end{equation}
where the tilde indicates the electric and magnetic components measured by the observer. In order to construct the spacetime Faraday tensor $F_{\mu\nu}$, we must project \eqref{faradaylocal} into spacetime.

\subsection{Characterization of the observer}\label{sec3.1}

In order to proceed, we need a set of tetrads associated with Schwarzschild coordinates, given by the line element
\begin{equation}
	ds^{2}=-f(r)dt^{2}+f(r)^{-1}dr^{2}+r^{2}d\Omega^{2}\,,
\end{equation}
with $f(r)=1-2M/r$, where $M$ is the Schwarzschild mass parameter, and adapted to some observer. Since we want to keep the interpretation of the components $\tilde{E}$ and $\tilde{B}$ as the electric and magnetic fields measured by a static observer, we require a set of tetrads adapted to such an observer. One set of tetrads that fulfills this requirement is
\begin{equation}\label{static}
e_{a\mu}=
\begingroup\scriptsize
\begin{pmatrix}
-f^{1/2} & 0 & 0 & 0 \\\\
0 & f^{-1/2}\cos\phi\sin\theta & r\cos\theta\cos\phi & -r\sin\theta\sin\phi \\\\
0 & f^{-1/2}\sin\phi\sin\theta & r\cos\theta\sin\phi & r\cos\phi\sin\theta \\\\
0 & f^{-1/2}\cos\theta & -r\sin\theta &0
\end{pmatrix}\,,
\endgroup
\end{equation}
where $e=det(e_{a}\,^{\mu})=r^{2}\sin\theta$.
One may check the consistency of the tetrads \eqref{static} as adapted to a static observer by noting that 
\begin{equation}
	u^{\mu}=g^{\mu\nu}e_{(0)\nu}=e_{(0)}{}^{\mu}=(f^{-1/2},0,0,0)\,.
\end{equation}
From \eqref{static} it is straightforward to verify that the non-zero components of the torsion tensor $T^{a}{}_{\mu\nu}$ are
\begin{align}
T^{(0)}{}_{01} &= -\frac{M/r^{2}}{\sqrt{1-2M/r}}, \nonumber\\
T^{(1)}{}_{12} &= \left(\frac{1}{\sqrt{1-2M/r}} - 1\right)\cos\theta \cos\phi, \nonumber\\
T^{(1)}{}_{13} &= -\left(\frac{1}{\sqrt{1-2M/r}} - 1\right)\sin\theta \sin\phi, \nonumber\\
T^{(2)}{}_{12} &= \left(\frac{1}{\sqrt{1-2M/r}} - 1\right)\cos\theta \sin\phi, \nonumber\\
T^{(2)}{}_{13} &= \left(\frac{1}{\sqrt{1-2M/r}} - 1\right)\sin\theta \cos\phi, \nonumber\\
T^{(3)}{}_{12} &= -\left(\frac{1}{\sqrt{1-2M/r}} - 1\right)\sin\theta\,.\label{torsion}
\end{align}
In the absence of a gravitational field, the torsion must vanish for a static observer, which can be seen from \eqref{torsion} by setting $M=0$.

From \eqref{torsion}, the acceleration tensor \eqref{mashhoon} can be evaluated as
\begin{align}
\phi_{(0)(1)} &= \frac{M}{r^{2} f^{1/2}}\,\cos\phi\,\sin\theta, \nonumber\\
\phi_{(0)(2)} &= \frac{M}{r^{2} f^{1/2}}\,\sin\theta\,\sin\phi, \nonumber\\
\phi_{(0)(3)} &= \frac{M}{r^{2} f^{1/2}}\,\cos\theta\,,\label{accelerationstaticcartesian}
\end{align}
while $\phi_{(i)(j)}=0$. By considering the radial direction in the tangent space $\hat{\mathbf r}=\cos{\phi}\sin{\theta}\,\hat{\mathbf x} + \sin{\phi}\sin{\theta}\,\hat{\mathbf y} + \cos{\theta}\,\hat{\mathbf z}$, we may write the inertial acceleration of the frame \eqref{accelerationstaticcartesian} as
\begin{equation}\label{inertial_acceleration}
\mathbf{a}_{i} = \frac{M}{r^{2} f^{1/2}}\,\hat{\mathbf r}\,.
\end{equation}
Such acceleration is the one required to maintain the frame in its static kinematical state, i.e., it is opposite to the gravitational acceleration $\mathbf{a}_{g}=-\frac{M}{r^{2} f^{1/2}}\,\hat{\mathbf r}$. Again, for consistency, one may verify that such acceleration vanishes for $M=0$, as expected.

Having fully characterized the observer and the associated tetrad field, we may now construct the electromagnetic field tensor in spacetime and derive the corresponding Maxwell equations.

\subsection{Generalization of Maxwell equations}\label{sec3.2}

Having characterized the observer, we are now ready to proceed with the computation of \(F_{\mu\nu}=e^{a}{}_{\mu}e^{b}{}_{\nu}F_{ab}\). From \eqref{faradaylocal} and \eqref{static}, we obtain
\begin{widetext}
\begin{equation}
	F_{\mu\nu}=
\begin{pmatrix}
0 & \tilde{E}_r & f^{1/2}r\tilde{E}_\theta & f^{1/2}r\sin{\theta}\tilde{E}_\phi \\\\
-\tilde{E}_r & 0 & -rf^{-1/2}\tilde{B}_\phi & rf^{-1/2}\sin\theta \tilde{B}_\theta \\\\
-f^{1/2}r\tilde{E}_\theta & rf^{-1/2}\tilde{B}_\phi & 0 & -r^2\sin\theta\tilde{B}_r \\\\
-rf^{1/2}\sin{\theta}\tilde{E}_\phi & -rf^{-1/2}\sin\theta\tilde{B}_\theta& r^2\sin\theta \tilde{B}_r &0
\end{pmatrix}\,,
\end{equation}
\end{widetext}
where we defined the components
\begin{align}
\tilde{A}_{r} &= \sin\theta\cos\phi\,\tilde{A}_{x}
              + \sin\theta\sin\phi\,\tilde{A}_{y}
              + \cos\theta\,\tilde{A}_{z}\,,\\
\tilde{A}_{\theta} &= \cos\theta\cos\phi\,\tilde{A}_{x}
                   + \cos\theta\sin\phi\,\tilde{A}_{y}
                   - \sin\theta\,\tilde{A}_{z}\,,\\ 
\tilde{A}_{\phi} &= -\sin\phi\,\tilde{A}_{x}
                 + \cos\phi\,\tilde{A}_{y}\,,
\end{align}
where \(\tilde{A}=\tilde{E},\tilde{B}\), i.e., the components of the electromagnetic field in the local basis written in spherical coordinates.

Maxwell equations in arbitrary coordinates read
\begin{equation}\label{maxwellspacetime}
	\mathring{\nabla}_{\mu}F^{\mu\nu}=J^{\nu}\,,\qquad 
	\mathring{\nabla}_{\mu}\mathcal{F}^{\mu\nu}=0\,,
\end{equation}
where \(\mathcal{F}^{\mu\nu}\) is the dual Faraday tensor. Since the electromagnetic field measured by an arbitrary observer is obtained by projection onto the local frame, it is natural to seek the corresponding form of Maxwell equations with one free Lorentz index.

Given the irrelevant role of the aether existence, Maxwell equations must transform covariantly under global Lorentz transformations. Since Lorentz transformations act on the frame, they act on the local indices of tensors, i.e., they must be of the form \(\Lambda^{a}{}_{b}\). As discussed in \cite{maluf2010electrodynamics}, coordinate and frame transformations may be disentangled, so that one may transform the frame independently of the coordinates, and vice-versa. Therefore, the electromagnetic field measured by an arbitrary observer should be described by quantities carrying local Lorentz indices.

Let us consider two frames \(S\) and \(\bar{S}\), related by a global Lorentz transformation \(\Lambda^{a}{}_{b}\). The local components of the vector potential transform as
\begin{equation}
	\bar{A}^{a}=\Lambda^{a}{}_{b}A^{b},
\end{equation}
and similarly for any projected tensorial quantity. In particular, if the electromagnetic field in the frame is described by \(F_{ab}\), then under a global Lorentz transformation one has
\begin{equation}
	\bar{F}_{ab}=\Lambda_{a}{}^{c}\Lambda_{b}{}^{d}F_{cd}.
\end{equation}
Thus Maxwell equations written with local Lorentz indices are covariant under global Lorentz transformations.

Following \cite{maluf2010electrodynamics}, we introduce the Levi-Civita covariant derivative in the local frame,
\begin{equation}
	D_{a}A_{b}=e_{a}{}^{\mu}\left(\partial_{\mu}A_{b}-\mathring{\omega}_{\mu}{}^{c}{}_{b}A_{c}\right),
\end{equation}
and define the Faraday tensor in the frame by
\begin{equation}
	F_{ab}=D_{a}A_{b}-D_{b}A_{a}.
\end{equation}
A direct computation shows that this quantity coincides with the tetrad projection of the usual spacetime Faraday tensor, namely,
\begin{equation}
	F_{ab}=e_{a}{}^{\mu}e_{b}{}^{\nu}F_{\mu\nu}.
\end{equation}
Therefore, the electromagnetic field measured in an arbitrary frame is obtained by projecting the spacetime tensor onto the tangent space, while the distinction between coordinate and frame transformations is preserved.

Maxwell equations in the local frame then follow from the Lagrangian density
\begin{equation}
	L=-\frac{1}{4}e\,F_{ab}F^{ab}-e\,A_{b}J^{b},
\end{equation}
whose variation yields
\begin{equation}\label{maxwelllocaldensitized}
	\partial_{\mu}(eF^{\mu b})+eF^{\mu c}\mathring{\omega}_{\mu}{}^{b}{}_{c}
	=eJ^{b}.
\end{equation}
The homogeneous equations, in turn, may be written as
\begin{equation}\label{maxwellhomlocal}
	D_{a}F_{bc}+D_{b}F_{ca}+D_{c}F_{ab}=0.
\end{equation}
Thus, both the homogeneous and inhomogeneous equations are completely equivalent to the standard spacetime Maxwell equations, but now written in terms of the field components measured in an arbitrary frame.

Expanding Eq.~\eqref{maxwelllocaldensitized} and using
\begin{equation}
	\partial_{\mu}e=e\,\mathring{\Gamma}^{\lambda}{}_{\mu\lambda},
\end{equation}
we obtain
\begin{equation}
	\partial_{\mu}F^{\mu b}
	+\mathring{\Gamma}^{\lambda}{}_{\mu\lambda}F^{\mu b}
	+\mathring{\omega}_{\mu}{}^{b}{}_{c}F^{\mu c}
	=
	J^{b}.
\end{equation}
Renaming dummy indices, the above equation may be rewritten as
\begin{equation}\label{inhomogeneous}
	\partial_{\mu}F^{\mu b}
	+\mathring{\Gamma}^{\mu}{}_{\lambda\mu}F^{\lambda b}
	+\mathring{\omega}_{\mu}{}^{b}{}_{c}F^{\mu c}
	=
	J^{b}.
\end{equation}

For the homogeneous equations, one may proceed analogously by introducing the dual Faraday tensor projected onto the frame,
\begin{equation}
	\mathcal{F}^{\mu b}=e^{b}{}_{\nu}\mathcal{F}^{\mu\nu}.
\end{equation}
Then the local form of the homogeneous equations reads
\begin{equation}\label{maxwelllocalhomdensitized}
	\partial_{\mu}(e\mathcal{F}^{\mu b})+e\mathcal{F}^{\mu c}\mathring{\omega}_{\mu}{}^{b}{}_{c}=0,
\end{equation}
or, equivalently,
\begin{equation}\label{homogeneous}
	\partial_{\mu}\mathcal{F}^{\mu b}
	+\mathring{\Gamma}^{\mu}{}_{\lambda\mu}\mathcal{F}^{\lambda b}
	+\mathring{\omega}_{\mu}{}^{b}{}_{c}\mathcal{F}^{\mu c}
	=0.
\end{equation}

These equations are obtained from the Levi-Civita covariant derivative and from the tetrad projection of the Faraday tensor. In this sense, \eqref{inhomogeneous} and \eqref{homogeneous} are the projected forms of the standard spacetime equations in \eqref{maxwellspacetime}.

The generalizations \eqref{inhomogeneous} and \eqref{homogeneous} allow us to treat the electromagnetic phenomenon for any frame of reference and for any gravitational field. The strength of this formulation is that we may analyze the electromagnetic fields in the same spacetime for different observers using the same coordinate system. This is a powerful tool for interpreting the phenomenon in the presence of gravity. In the next two sections we investigate the electromagnetic field in Schwarzschild spacetime in two distinct frames of reference.

\section{Maxwell equations for a static observer}\label{sec4}

In order to compute \eqref{inhomogeneous} and \eqref{homogeneous}, we must evaluate the non-zero components of the Levi-Civita connection \eqref{levi}. They read
\begin{align}
\mathring\omega_{0(0)(1)}&=-\frac{M}{r^2}\cos{\phi}\sin{\theta}\,,\nonumber\\
\mathring\omega_{0(0)(2)}&=-\frac{M}{r^2}\sin{\phi}\sin{\theta}\,,\nonumber\\
\mathring\omega_{0(0)(3)}&=-\frac{M}{r^2}\cos{\theta}\,,\nonumber\\
\mathring\omega_{2(2)(3)}&=-\frac{-2M+r-f^{1/2}r}{f^{1/2}r}\cos\phi\,,\nonumber\\
\mathring\omega_{2(3)(1)}&=-\frac{2M+(-1+f^{1/2})r}{f^{1/2}r}\cos\phi\,,\nonumber\\
\mathring\omega_{2(3)(2)}&=-\frac{2M+(-1+f^{1/2})r}{f^{1/2}r}\sin\phi\,,\nonumber\\
\mathring\omega_{3(1)(2)}&=-\frac{2M+(-1+f^{1/2})r}{f^{1/2}r}\sin^2\theta\,,\nonumber\\
\mathring\omega_{3(1)(3)}&=-\frac{2M+(-1+f^{1/2})r}{f^{1/2}r}\cos\theta\sin\theta\sin\phi\,,\nonumber\\
\mathring\omega_{3(2)(1)}&=-\frac{-2M+r-f^{1/2}r}{f^{1/2}r}\sin^2\theta\,,\nonumber\\
\mathring\omega_{3(2)(3)}&=\frac{2M+(-1+f^{1/2})r}{f^{1/2}r}\cos\theta\cos\phi\sin\theta\,,\nonumber\\
\mathring\omega_{3(3)(1)}&=-(-1+f^{1/2})\cos\theta\sin\theta\sin\phi\,,\nonumber\\
\mathring\omega_{3(3)(2)}&=(-1+f^{1/2})\cos\theta\cos\phi\sin\theta\,.\nonumber
\end{align}

In order to simplify the calculations, it is convenient to write \eqref{inhomogeneous} in the form
\begin{equation}\label{inhomogeneousdet}
	\partial_{\mu}(eF^{\mu b})+e\,\mathring{\omega}_{\mu}{}^{b}{}_{c}\,F^{\mu c}=eJ^{b}\,,
\end{equation}
and \eqref{homogeneous} as
\begin{equation}\label{homogeneousdet}
	\partial_{\mu}(e\mathcal{F}^{\mu b})+e\,\mathring{\omega}_{\mu}{}^{b}{}_{c}\,\mathcal{F}^{\mu c}=0\,.
\end{equation}

The four-current in the local basis is
\begin{equation}\label{4-current_static}
J^{b}=(\tilde{\rho},\tilde{J}^{(1)},\tilde{J}^{(2)},\tilde{J}^{(3)})\,.
\end{equation}
Its spatial part may be written in Cartesian form,
\begin{equation}
(\tilde{J}^{(1)},\tilde{J}^{(2)},\tilde{J}^{(3)})=(\tilde{J}_{x},\tilde{J}_{y},\tilde{J}_{z})\,.
\end{equation}
This is the four-current measured by the observer in its local frame. In particular, $\tilde{\rho}$ denotes the charge density measured by that observer. In order to proceed further, we define
\begin{align}
\tilde{J}_{r} &= \sin\theta\cos\phi\,\tilde{J}_{x}
              + \sin\theta\sin\phi\,\tilde{J}_{y}
              + \cos\theta\,\tilde{J}_{z}\,,\nonumber\\
\tilde{J}_{\theta} &= \cos\theta\cos\phi\,\tilde{J}_{x}
                   + \cos\theta\sin\phi\,\tilde{J}_{y}
                   - \sin\theta\,\tilde{J}_{z}\,,\nonumber\\
\tilde{J}_{\phi} &= -\sin\phi\,\tilde{J}_{x}
                 + \cos\phi\,\tilde{J}_{y}\,,\label{currentspherical}
\end{align}
as the spherical components of the measured four-current in the local frame.

We now have the necessary ingredients to proceed. In the following subsections, the inhomogeneous and homogeneous equations are evaluated separately.

\subsection{Inhomogeneous equations for a static observer}\label{sec4.1}

By choosing $b=(0)$ in \eqref{inhomogeneousdet}, one obtains, after straightforward calculations, Gauss's law,
\begin{equation}\label{Gauss_static}
\frac{f^{1/2}}{r^2}\partial_r(r^2\tilde{E}_r)+\frac{1}{r\sin\theta}\partial_\theta(\sin\theta\tilde{E}_\theta)+\frac{1}{r\sin\theta}\partial_\phi(\tilde{E}_\phi)=\tilde{\rho}\,.
\end{equation}
We may see that the gravitational correction appears only in the radial contribution through the factor $f^{1/2}$, while the angular sector preserves its usual spherical form. This reflects the fact that the Schwarzschild geometry modifies the radial structure of spacetime while preserving its angular symmetry.

For Amp\`ere-Maxwell law, we choose $b=(i)$ in \eqref{inhomogeneousdet}, obtaining a set of lengthy equations for the Cartesian components of the current. They read
\begin{widetext}
\begin{align}
\frac{1}{f^{1/2} r^2} \Big[
&\cos\phi \Big\{
(M + (-1 + f^{1/2}) r)\cos\theta \,\tilde{B}_\phi
+ f^{1/2} r \Big( \cot\theta \,\partial_\phi \tilde{B}_r
- \partial_\phi \tilde{B}_\theta
+ \sin\theta \,\partial_\theta \tilde{B}_\phi \Big) \nonumber \\
&\qquad - r \Big[ (-2M + r)\cos\theta \,\partial_r \tilde{B}_\phi
+ r \big( \sin\theta \,\partial_t \tilde{E}_r
+ \cos\theta \,\partial_t \tilde{E}_\theta \big) \Big]
\Big\} \nonumber \\
&+ \sin\phi \Big\{
(M - r)\tilde{B}_\theta
+ r \Big[ (2M - r)\partial_r \tilde{B}_\theta
+ r \,\partial_t \tilde{E}_\phi \Big]
\Big\}
\Big]
= \tilde{J}^{(1)}\,,\label{AMx}
\end{align}

\begin{align}
\frac{1}{f^{1/2} r^2} \Big[
&\sin\phi \Big\{
f^{1/2} r \Big(
\cos\theta \,\tilde{B}_\phi
+ \cot\theta \,\partial_\phi \tilde{B}_r
- \partial_\phi \tilde{B}_\theta
+ \sin\theta \,\partial_\theta \tilde{B}_\phi
\Big) \nonumber \\
&\qquad + (M - r)\cos\theta \,\tilde{B}_\phi
- r \Big[ (-2M + r)\cos\theta \,\partial_r \tilde{B}_\phi
+ r \big( \sin\theta \,\partial_t \tilde{E}_r
+ \cos\theta \,\partial_t \tilde{E}_\theta \big) \Big]
\Big\} \nonumber \\
&+ \cos\phi \Big\{
(-M + r)\tilde{B}_\phi
+ r \Big[ (-2M + r)\partial_r \tilde{B}_\theta
- r \,\partial_t \tilde{E}_\phi \Big]
\Big\}
\Big]
= \tilde{J}^{(2)}\,,\label{AMy}
\end{align}

\begin{align}
\frac{1}{f^{1/2} r^2} \Big[
&\tilde{B}_\phi \Big(
f^{1/2} r \cos\theta \cot\theta
+ (-M + r)\sin\theta
\Big) \nonumber \\
&+ r \Big\{
f^{1/2} \Big(
- \partial_\phi \tilde{B}_r
- \cot\theta \,\partial_\phi \tilde{B}_\theta
+ \cos\theta \,\partial_\theta \tilde{B}_\phi
\Big) \nonumber \\
&\qquad + (-2M + r)\sin\theta \,\partial_r \tilde{B}_\phi
- r \cos\theta \,\partial_t \tilde{E}_r
+ r \sin\theta \,\partial_t \tilde{E}_\theta
\Big\}
\Big]
= \tilde{J}^{(3)}\,.\label{AMz}
\end{align}
\end{widetext}

Although the field components are naturally obtained in the local Cartesian basis associated with the tetrads, it is more convenient to recombine them into spherical components, since this form better reflects the symmetry of Schwarzschild spacetime.

Hence, the above equations can be combined in such a way that the final result acquires a simple spherical form. Indeed, by multiplying \eqref{AMx} by $\sin\theta\cos\phi$, \eqref{AMy} by $\sin\theta\sin\phi$, and \eqref{AMz} by $\cos\theta$, and then summing the resulting expressions, one obtains
\begin{equation}\label{AMr}
\frac{1}{r\sin\theta}\left(\partial_\theta(\sin\theta\tilde{B}_\phi)-\partial_\phi(\tilde{B}_\theta)\right)=\tilde{J}_{r}+f^{-1/2}\partial_t\tilde{E}_r\,.
\end{equation}

Proceeding similarly, by multiplying \eqref{AMx} by $\cos\theta\cos\phi$, \eqref{AMy} by $\cos\theta\sin\phi$, and \eqref{AMz} by $-\sin\theta$, one obtains
\begin{equation}\label{AMtheta}
\frac{1}{r}\left[\frac{1}{\sin\theta}\partial_\phi(\tilde{B}_r)-\partial_r(rf^{1/2}\tilde{B}_\phi)\right]=\tilde{J}_{\theta}+f^{-1/2}\partial_t\tilde{E}_\theta\,.
\end{equation}

Finally, by multiplying \eqref{AMx} by $-\sin\phi$ and \eqref{AMy} by $\cos\phi$, one obtains
\begin{equation}\label{AMphi}
\frac{1}{r}\partial_r(rf^{1/2}\tilde{B}_\theta)-\frac{1}{r}\partial_\theta(\tilde{B}_r)=\tilde{J}_{\phi}+f^{-1/2}\partial_t\tilde{E}_\phi\,.
\end{equation}

In deriving the above equations, the relations \eqref{currentspherical} have been used. Therefore, Eqs.~\eqref{AMr}--\eqref{AMphi} correspond to Amp\`ere-Maxwell law written in terms of the electromagnetic fields measured by the static observer.

It is interesting to note that the gravitational field modifies the radial magnetic contributions through the factor $f^{1/2}$, while the displacement-current term acquires the factor $f^{-1/2}$. Hence, even though the final equations retain the familiar spherical structure, the gravitational field affects differently the magnetic and electric sectors of Amp\`ere-Maxwell law.

The recovery of the Minkowski expressions in the limit $f(r)=1$ provides a consistency check of the formalism and of the interpretation of the projected fields as the physical fields measured by the observer.

Thus, the inhomogeneous Maxwell equations in Schwarzschild spacetime, as measured by a static observer, are given by Gauss's law together with Eqs.~\eqref{AMr}--\eqref{AMphi}.

\subsection{Homogeneous equations for a static observer}\label{sec4.2}

We now turn to the homogeneous equations \eqref{homogeneousdet}. The procedure is entirely analogous to that employed in the previous subsection. For this reason, and in order to avoid unnecessary repetition, we present directly the final expressions in spherical components.

By choosing $b=(0)$ in \eqref{homogeneousdet}, we obtain magnetic Gauss's law,
\begin{equation}\label{GaussB_static}
\frac{f^{1/2}}{r^2}\partial_r(r^2\tilde{B}_r)+\frac{1}{r\sin\theta}\partial_\theta(\sin\theta\tilde{B}_\theta)+\frac{1}{r\sin\theta}\partial_\phi(\tilde{B}_\phi)=0\,.
\end{equation}
As in the electric case, the gravitational correction appears only in the radial contribution, whereas the angular sector preserves its usual spherical form.

For Faraday's law, we choose $b=(i)$ in \eqref{homogeneousdet}. After carrying out the same projection procedure used in the inhomogeneous case, we obtain the spherical components of the equation. The radial component is
\begin{equation}\label{Faradayr}
\frac{1}{r\sin\theta}\left[\partial_\theta(\sin\theta\tilde{E}_\phi)-\partial_\phi(\tilde{E}_\theta)\right]=-f^{-1/2}\partial_t\tilde{B}_r\,.
\end{equation}

Proceeding analogously, we obtain the polar component
\begin{equation}\label{Faradaytheta}
\frac{1}{r}\left[\frac{1}{\sin\theta}\partial_\phi(\tilde{E}_r)-\partial_r(rf^{1/2}\tilde{E}_\phi)\right]=-f^{-1/2}\partial_t\tilde{B}_\theta\,,
\end{equation}
and the azimuthal component
\begin{equation}\label{Faradayphi}
\frac{1}{r}\partial_r(rf^{1/2}\tilde{E}_\theta)-\frac{1}{r}\partial_\theta(\tilde{E}_r)=-f^{-1/2}\partial_t\tilde{B}_\phi\,.
\end{equation}

Therefore, the homogeneous equations preserve the same general spherical structure found in flat spacetime, but with the gravitational field modifying the radial and temporal sectors through the factors $f^{1/2}$ and $f^{-1/2}$, respectively. This behavior is entirely analogous to that found in the inhomogeneous equations.

Once again, in the limit $f(r)=1$, the usual Minkowski expressions in spherical coordinates are recovered.

\section{Maxwell equations for a radially free-falling observer}\label{sec5}

In order to compute Maxwell equations for a radially free-falling frame, we must first construct a set of tetrads $\bar{e}_{a\mu}$ adapted to a radially geodesic observer, whose four-velocity reads
\begin{equation}\label{u_freefalling}
	u^{\mu}=-\bar{e}^{(0)\mu}=(f^{-1},-\sqrt{1-f},0,0)\,.
\end{equation}
Hence, we begin with a set of inverse tetrads of the form
\begin{equation}
	\bar{e}^{a\mu}=
	\begingroup\scriptsize
\left(
\begin{array}{cccc}
-\dfrac{1}{f} & \sqrt{1-f} & 0 & 0 \\[0.8em]
-A\cos\phi\,\sin\theta & B\cos\phi\,\sin\theta
& \dfrac{\cos\theta\,\cos\phi}{r}
& -\dfrac{\csc\theta\,\sin\phi}{r} \\[0.8em]
-A\sin\theta\,\sin\phi & B\sin\theta\,\sin\phi
& \dfrac{\cos\theta\,\sin\phi}{r}
& \dfrac{\cos\phi\,\csc\theta}{r} \\[0.8em]
-A\cos\theta & B\cos\theta
& -\dfrac{\sin\theta}{r}
& 0
\end{array}
\right)\,,
\endgroup
\end{equation}
where $A$ and $B$ are functions to be determined.

By imposing the orthonormality relation \eqref{spacetimemetric}, we obtain
\begin{equation}
	B=1\,,\qquad A=f^{-1}\sqrt{1-f}\,.
\end{equation}
Therefore, the covariant tetrads read
\begin{equation}\label{tetrads_freefalling}
\bar{e}_{a\mu}=
\begingroup\scriptsize
\left(
\begin{array}{cccc}
-1 & -\dfrac{\sqrt{1-f}}{f} & 0 & 0 \\[0.8em]
\sqrt{1-f}\cos\phi\sin\theta
&
\dfrac{\cos\phi\sin\theta}{f}
&
r\cos\theta\cos\phi
&
-r\sin\theta\sin\phi
\\[0.8em]
\sqrt{1-f}\sin\theta\sin\phi
&
\dfrac{\sin\theta\sin\phi}{f}
&
r\cos\theta\sin\phi
&
r\cos\phi\sin\theta
\\[0.8em]
\sqrt{1-f}\cos\theta
&
\dfrac{\cos\theta}{f}
&
-r\sin\theta
&
0
\end{array}
\right)\,.
\endgroup
\end{equation}
The determinant of this tetrad coincides with that of the static frame, namely $\bar{e}=e=r^{2}\sin\theta$.

The non-zero components of the torsion tensor associated with \eqref{tetrads_freefalling} are
\begin{align}
T_{001} &= \frac{f'}{2},\nonumber
&
T_{110} &= -\frac{f'}{2\sqrt{1-f}\,f},\nonumber
\\[0.5em]
T_{202} &= -r\sqrt{1-f},\nonumber
&
T_{212} &= r-\frac{r}{f},\nonumber
\\[0.5em]
T_{303} &= -r\sqrt{1-f}\,\sin^{2}\theta,\nonumber
&
T_{331} &= -\frac{r(-1+f)\sin^{2}\theta}{f}\,,
\end{align}
where $f'=df/dr$.

Since the frame is adapted to a radially geodesic observer, its translational inertial acceleration must vanish. Indeed, the acceleration tensor is identically zero,
\begin{equation}
\phi_{ab}=0\,,
\end{equation}
showing that the frame is freely falling and non-rotating in the sense of the Mashhoon tensor. This is the fundamental difference with respect to the static frame considered in Sec.~\ref{sec4}, where a non-vanishing inertial acceleration is required in order to keep the observer at rest in Schwarzschild spacetime.

As a consistency check, in the limit $M\to 0$ one has $f\to 1$, and the tetrads \eqref{tetrads_freefalling} reduce to those of a static observer in Minkowski spacetime. In the same limit, the torsion components vanish, as expected.

The non-zero components of the Levi-Civita connection in the free-falling frame are
\begingroup\small
\begin{alignat}{2}
\mathring{\omega}_{0(1)(0)}
&= -\frac{1}{2}\cos\phi\,\sin\theta\, f',
\qquad&
\mathring{\omega}_{0(2)(0)}
&= -\frac{1}{2}\sin\theta\,\sin\phi\, f',
\nonumber\\[0.5em]
\mathring{\omega}_{0(3)(0)}
&= -\frac{1}{2}\cos\theta\, f',
&
\mathring{\omega}_{1(1)(0)}
&= -\frac{\cos\phi\,\sin\theta\, f'}{2\sqrt{1-f}\,f},
\nonumber\\[0.5em]
\mathring{\omega}_{1(2)(0)}
&= -\frac{\sin\theta\,\sin\phi\, f'}{2\sqrt{1-f}\,f},
&
\mathring{\omega}_{1(3)(0)}
&= -\frac{\cos\theta\, f'}{2\sqrt{1-f}\,f},
\nonumber\\[0.5em]
\mathring{\omega}_{2(1)(0)}
&= \cos\theta\,\cos\phi\,\sqrt{1-f},
&
\mathring{\omega}_{2(2)(0)}
&= \cos\theta\,\sqrt{1-f}\,\sin\phi,
\nonumber\\[0.5em]
\mathring{\omega}_{2(3)(0)}
&= -\sqrt{1-f}\,\sin\theta,
&
\mathring{\omega}_{3(1)(0)}
&= -\sqrt{1-f}\,\sin\theta\,\sin\phi,
\nonumber\\[0.5em]
\mathring{\omega}_{3(2)(0)}
&= \cos\phi\,\sqrt{1-f}\,\sin\theta\,.
\nonumber
\end{alignat}
\endgroup

We also need the four-current measured in the free-falling frame. Since we wish to keep explicit the components measured in the static frame, we first project the four-current \eqref{4-current_static} into spacetime as
\begin{equation}
	J^{\mu}=e_{b}{}^{\mu}\tilde{J}^{b}\,,
\end{equation}
and then project $J^{\mu}$ into the free-falling frame:
\begin{equation}
	\bar{J}^{a}=\bar{e}^{a}{}_{\mu}J^{\mu}=\Lambda^{a}{}_{b}\tilde{J}^{b}\,,
\end{equation}
where we have defined the local Lorentz transformation
\begin{equation}\label{lorentz_transformation}
	\Lambda^{a}{}_{b}=\bar{e}^{a}{}_{\mu}e_{b}{}^{\mu}\,.
\end{equation}
From \eqref{static} and \eqref{tetrads_freefalling}, we obtain
\begin{widetext}
\begin{equation}\label{Lambda_radial_boost}
	\Lambda^{a}{}_{b}
	=
	\begin{pmatrix}
		\gamma & \alpha n_x & \alpha n_y & \alpha n_z\\
		\alpha n_x & 1+(\gamma-1)n_x^2 & (\gamma-1)n_x n_y & (\gamma-1)n_x n_z\\
		\alpha n_y & (\gamma-1)n_y n_x & 1+(\gamma-1)n_y^2 & (\gamma-1)n_y n_z\\
		\alpha n_z & (\gamma-1)n_z n_x & (\gamma-1)n_z n_y & 1+(\gamma-1)n_z^2
	\end{pmatrix},
\end{equation}
\end{widetext}
where
\begin{equation}\label{boost_parameters}
	\gamma=\frac{1}{\sqrt{f(r)}}\,,
	\qquad
	\alpha=\sqrt{\frac{1}{f(r)}-1}\,,
\end{equation}
and
\begin{equation}\label{radial_unit_vector}
	n^{(i)}=
	\left(
		\sin\theta\cos\phi,
		\sin\theta\sin\phi,
		\cos\theta
	\right).
\end{equation}
Thus, from \eqref{lorentz_transformation}, the components of the current transform as
\begin{subequations}\label{current_boost_cartesian}
	\begin{align}
	\bar{J}^{(0)}
	&=
	\gamma \tilde{J}^{(0)}
	+
	\alpha n_{(i)}\tilde{J}^{(i)},
	\\
	\bar{J}^{(i)}
	&=
	\alpha n^{(i)} \tilde{J}^{(0)}
	+
	\left[
		\delta^{i}{}_{j}
		+
		(\gamma-1)n^{(i)} n_{(j)}
	\right]\tilde{J}^{(j)}.
\end{align}
\end{subequations}

In spherical components, this reduces to
\begin{subequations}\label{current_boost_spherical}
\begin{align}
\bar{J}^{(0)}
&=
\gamma \tilde{\rho} + \alpha \tilde{J}_{r},
\\[0.5em]
\bar{J}^{(r)}
&=
\alpha \tilde{\rho} + \gamma \tilde{J}_{r},
\\[0.5em]
\bar{J}^{(\theta)}
&=
\tilde{J}_{\theta},
\\[0.5em]
\bar{J}^{(\phi)}
&=
\tilde{J}_{\phi}.
\end{align}
\end{subequations}
As expected, only the temporal and radial components are mixed, since the free-falling frame is related to the static one by a radial boost.

With the above results, we may proceed in complete analogy with Sec.~\ref{sec4} and evaluate Maxwell equations in the next two subsections.

\subsection{Inhomogeneous equations for a free-falling observer}\label{sec5.1}

We now evaluate the inhomogeneous Maxwell equations for the radially free-falling observer. The procedure is the same as that employed for the static frame. However, since the free-falling tetrad is related to the static tetrad by a radial boost, the temporal-radial projections of Maxwell equations mix the electric and magnetic sectors, while the angular projections retain the same structure found for the static observer. This feature is already visible in Gauss's law, where, by choosing $b=(0)$ in \eqref{inhomogeneousdet}, we obtain
\begin{align}
&\frac{1}{r^2}\partial_r\!\left(r^2\tilde{E}_r\right)
+\frac{1}{r f^{1/2}\sin\theta}
\left[
\partial_\theta\!\left(\sin\theta\,\tilde{E}_\theta\right)
+\partial_\phi\tilde{E}_\phi
\right] \nonumber\\
&
+\frac{\sqrt{1-f}}{r f^{1/2}\sin\theta}
\left[
\partial_\theta\!\left(\sin\theta\,\tilde{B}_\phi\right)
-\partial_\phi\tilde{B}_\theta
\right] \nonumber\\
&
-\frac{\sqrt{1-f}}{f}\,\partial_t\tilde{E}_r
=
\gamma \tilde{\rho}+\alpha\tilde{J}_{r}\,.
\label{Gauss_freefalling}
\end{align}
Unlike the static case, Gauss's law no longer contains only the divergence of the electric field. The radial boost of the frame produces additional terms involving the angular magnetic components and the time derivative of the radial electric field. The source term also changes accordingly, since the charge density measured by the free-falling observer is $\bar{J}^{(0)}=\gamma \tilde{\rho}+\alpha\tilde{J}_{r}$.

For the spatial components, $b=(i)$, one obtains Amp\`ere-Maxwell law. The radial component reads
\begin{align}
&\frac{\sqrt{1-f}}{f\,r^{2}\sin\theta}
\left\{
f\sin\theta\,\partial_r\!\left(r^2\tilde{E}_r\right)
+r f^{1/2}
\left[
\partial_\theta\!\left(\sin\theta\,\tilde{E}_\theta\right)\right.\right.\nonumber\\
&+\left.\left.
\partial_\phi \tilde{E}_\phi
\right]
\right\}
+\frac{1}{f^{1/2} r \sin{\theta} }
\left[
\partial_{\theta}(\sin{\theta}\,\tilde{B}_{\phi}) - \partial_{\phi}\tilde{B}_{\theta}
\right]\nonumber\\
&-\frac{1}{f}\partial_t\tilde{E}_r
=
\alpha\tilde{\rho}+\gamma\tilde{J}_{r}\,.
\label{AMr_freefalling}
\end{align}

This equation corresponds to the radial projection of Amp\`ere-Maxwell law in the free-falling frame. Its source is the radial current measured by that observer, namely $\bar{J}^{(r)}=\alpha\tilde{\rho}+\gamma\tilde{J}_{r}$.

The polar component is
\begin{small}
\begin{equation}\label{AMtheta_freefalling}
\frac{1}{r}
\left[
\frac{1}{\sin\theta}\partial_\phi\tilde{B}_r
-\partial_r\!\left(r f^{1/2}\tilde{B}_\phi\right)
\right]
-\frac{1}{f^{1/2}}\partial_t\tilde{E}_\theta
=
\tilde{J}_{\theta}\,.
\end{equation}
\end{small}
while the azimuthal component is
\begin{small}
\begin{equation}\label{AMphi_freefalling}
\frac{1}{r}
\left[
\partial_r\!\left(r f^{1/2}\tilde{B}_\theta\right)
-\partial_\theta\tilde{B}_r
\right]
-\frac{1}{f^{1/2}}\partial_t\tilde{E}_\phi
=
\tilde{J}_{\phi}\,.
\end{equation}
\end{small}
The angular components coincide with the corresponding equations for the static observer, written with the time-derivative terms on the left-hand side. This happens because the radial boost does not mix the angular components of the current nor the angular projections of the inhomogeneous equations. Indeed, as shown in \eqref{current_boost_spherical}, $\bar{J}^{(\theta)}=\tilde{J}_{\theta}$ and $\bar{J}^{(\phi)}=\tilde{J}_{\phi}$.

Equations \eqref{Gauss_freefalling}--\eqref{AMphi_freefalling} are the inhomogeneous Maxwell equations measured by a radially free-falling observer. In the limit $f\to 1$, one has $\alpha\to 0$ and $\gamma\to 1$, so that the equations reduce to the usual Minkowski expressions in spherical coordinates.

\subsection{Homogeneous equations for a free-falling observer}\label{sec5.2}

We now turn to the homogeneous equations. As before, the intermediate Cartesian components are omitted, since the procedure is entirely analogous to that used in the previous subsection. The final spherical equations are sufficient to display the physical content of the result.

By choosing $b=(0)$ in \eqref{homogeneousdet}, one obtains the magnetic Gauss's law,
\begin{align}
&\frac{1}{r^2}\partial_r\!\left(r^2\tilde{B}_r\right)
+\frac{1}{r f^{1/2}\sin\theta}
\left[
\partial_\theta\!\left(\sin\theta\,\tilde{B}_\theta\right)
+\partial_\phi\tilde{B}_\phi
\right] \nonumber\\
&
-\frac{\sqrt{1-f}}{r f^{1/2}\sin\theta}
\left[
\partial_\theta\!\left(\sin\theta\,\tilde{E}_\phi\right)
-\partial_\phi\tilde{E}_\theta
\right] \nonumber\\
&
-\frac{\sqrt{1-f}}{f}\,\partial_t\tilde{B}_r
=0\,.
\label{GaussB_freefalling}
\end{align}
As in the electric Gauss's law, the radial boost introduces a coupling with the angular sector of the other field. Thus, the condition of vanishing magnetic charge is expressed in the free-falling frame by a combination of the magnetic divergence, angular electric terms and the time derivative of $\tilde{B}_r$.

For the spatial components, one obtains Faraday's law. The radial component is
\begin{align}
&\frac{1}{r f^{1/2}\sin\theta}
\left[
\partial_\theta\!\left(\sin\theta\,\tilde{E}_\phi\right)
-\partial_\phi\tilde{E}_\theta
\right]
-\sqrt{1-f}\nonumber\\
&\times\left[
\frac{1}{r^2}\partial_r\!\left(r^2\tilde{B}_r\right)
+\frac{1}{f^{1/2}r\sin\theta}
\left(
\partial_\theta\!\left(\sin\theta\,\tilde{B}_\theta\right)
+\partial_\phi\tilde{B}_\phi
\right)
\right] \nonumber\\
&=-\frac{1}{f}\partial_t\tilde{B}_r\,.
\label{Faradayr_freefalling}
\end{align}

The polar component is
\begin{small}
\begin{equation}\label{Faradaytheta_freefalling}
\frac{1}{r}
\left[
\frac{1}{\sin\theta}\partial_\phi\tilde{E}_r
-\partial_r\!\left(r f^{1/2}\tilde{E}_\phi\right)
\right]
+\frac{1}{f^{1/2}}\partial_t\tilde{B}_\theta = 0\,.
\end{equation}
\end{small}
and the azimuthal component is
\begin{small}
\begin{equation}\label{Faradayphi_freefalling}
\frac{1}{r}
\left[
\partial_r\!\left(r f^{1/2}\tilde{E}_\theta\right)
-\partial_\theta\tilde{E}_r
\right]
+\frac{1}{f^{1/2}}\partial_t\tilde{B}_\phi =0\,.
\end{equation}
\end{small}

Eqs.~\eqref{GaussB_freefalling}--\eqref{Faradayphi_freefalling} show that the homogeneous sector also acquires mixed electric and magnetic contributions in the temporal-radial projections of the radially free-falling frame. The angular components of Faraday's law, however, coincide with those obtained for the static observer, again written with the time-derivative terms on the left-hand side. This mixing does not indicate the presence of magnetic charges; it is a consequence of the observer dependence of the measured electric and magnetic fields. In the limit $f\to 1$, the boost disappears and the standard flat spacetime Faraday's law and magnetic Gauss's law are recovered.

\section{Analysis}\label{sec6}

The equations obtained in the previous sections show that the tensorial form of Maxwell equations is preserved, whereas the separation into electric field, magnetic field, charge density and current density depends on the observer. In the static frame, the effect of the Schwarzschild geometry appears mainly through radial and temporal metric factors. In the radially free-falling frame, however, the equations also contain the kinematical effect of a local radial boost with respect to the static observer.

\subsection{Static observers and geometrical corrections}

The equations obtained for the static frame provide the simplest
reference point for the observer-dependent formulation. In this case, the
observer is adapted to the timelike Killing direction of Schwarzschild
spacetime and remains at fixed values of the spatial coordinates
$(r,\theta,\phi)$. Therefore, the corresponding measurements are made by
an observer which is at rest with respect to the central source, but not in
free fall.

The physical meaning of the correction factors appearing in Sec.~\ref{sec4}
can be understood directly from the local geometry. For the Schwarzschild
line element, the proper time measured by a static observer satisfies
\begin{equation}
	d\tau_{\rm stat}=f^{1/2}\,dt ,
\end{equation}
whereas the radial proper distance on a hypersurface $t=\mathrm{const.}$ is
given by
\begin{equation}
	d\ell_r=f^{-1/2}\,dr .
\end{equation}
Hence, the natural temporal and radial derivatives measured in the local
static frame are
\begin{equation}
	\partial_{\hat{0}}
	\equiv
	\frac{\partial}{\partial \tau_{\rm stat}}
	=
	f^{-1/2}\partial_t ,
	\qquad
	\partial_{\hat{r}}
	\equiv
	\frac{\partial}{\partial \ell_r}
	=
	f^{1/2}\partial_r .
\end{equation}
The factors $f^{-1/2}$ multiplying the time derivatives in
Eqs.~\eqref{AMr}--\eqref{AMphi} and
Eqs.~\eqref{Faradayr}--\eqref{Faradayphi} therefore have a direct
interpretation: they convert coordinate-time variations into variations
with respect to the proper time of the static observer.

The same interpretation applies to the radial sector of the divergence
terms. For instance, the radial contribution to the divergence of a vector
field $X_r$ measured in the static frame can be written as
\begin{equation}
	\frac{f^{1/2}}{r^2}\partial_r\left(r^2 X_r\right)
	=
	\frac{1}{r^2}\partial_{\hat r}\left(r^2 X_r\right).
\end{equation}
Thus, the factor $f^{1/2}$ does not represent a coupling between electric
and magnetic sectors. Rather, it expresses the fact that the radial
coordinate $r$ is not the proper radial distance measured by the static
observer.

A similar interpretation holds for the radial part of the curl terms.
Whenever a transverse component $X_A$, with $A=\theta,\phi$, appears in the
combination
\begin{equation}
	\frac{1}{r}\partial_r\left(r f^{1/2} X_A\right),
\end{equation}
one may rewrite it as
\begin{equation}
	\frac{1}{r}\partial_r\left(r f^{1/2} X_A\right)
	=
	\frac{1}{r}\partial_{\hat r}\left(r X_A\right)
	+
	\frac{f'}{2f^{1/2}}X_A .
\end{equation}
This form separates two effects. The first term is the usual radial
variation written in terms of proper radial distance. The second one is a
genuine geometrical correction associated with the radial variation of the
lapse function. In the unexpanded form used in
Eqs.~\eqref{AMtheta}--\eqref{AMphi} and
Eqs.~\eqref{Faradaytheta}--\eqref{Faradayphi}, this contribution is encoded
inside the derivative of $f^{1/2}$.

The angular sector has a different character. The angular terms appearing
in the static equations keep the usual spherical structure, with factors
$r$ and $r\sin\theta$ associated with the areal radius of the two-spheres
$r=\mathrm{const.}$ This reflects the spherical symmetry of Schwarzschild
spacetime. The gravitational field modifies the radial and temporal
sectors, but it does not introduce an angular anisotropy in the local
measurements of the static observer.

It is important to distinguish these geometrical corrections from the
electric-magnetic mixing that will appear in the temporal-radial projections
of the free-falling frame. In the static frame, Gauss's law involves only
the electric field and the electric charge density, while the magnetic
Gauss's law involves only the magnetic field. Similarly, Amp\`ere-Maxwell
law has the same structural content as in flat spacetime: a magnetic curl
term, an electric displacement-current term, and the electric current
source. Faraday's law preserves the complementary structure involving the
electric curl and the time derivative of the magnetic field. Thus, the
static frame deforms the differential operators, but it does not mix the
electric and magnetic sectors.

The price for this simple separation is that the static observer is not
geodesic. Its four-acceleration is non-vanishing. In Schwarzschild
coordinates, for an observer with
$u^\mu_{\rm stat}=f^{-1/2}\delta^\mu_t$, one finds the inertial acceleration given by Eq.~\eqref{inertial_acceleration}. Therefore, the static frame is physically maintained by an outward
acceleration which compensates the gravitational attraction. This
acceleration diverges as $r$ approaches the Schwarzschild radius, signalling
the well-known breakdown of the static frame at the horizon.

The static equations should therefore be interpreted as Maxwell equations
measured by accelerated observers which remain at rest with respect to the
Schwarzschild source. Their main role is to provide a reference
decomposition in which the gravitational corrections appear as metric
factors associated with proper time, proper radial distance, and the radial
variation of the lapse. In contrast, the free-falling frame considered
below introduces an additional effect: a local radial boost relative to the
static observers, which mixes charge and radial current, and mixes electric
and magnetic sectors in the temporal-radial projections of Maxwell
equations.

\subsection{Free fall as a radial boost}

The radially free-falling frame constructed in Sec.~\ref{sec5} is not
obtained from the static frame by a mere change of spatial coordinates. It
corresponds to a different family of physical observers. While the static
observer remains at fixed values of $(r,\theta,\phi)$, the observer
described by Eq.~\eqref{u_freefalling} follows a radial geodesic. Therefore,
the difference between both frames is local and kinematical: at each
spacetime point, the free-falling tetrad is related to the static one by a
Lorentz boost along the radial direction.

This interpretation follows directly from Eq.~\eqref{Lambda_radial_boost}.
Indeed, using the boost parameters defined in Eq.~\eqref{boost_parameters},
one may introduce the local relative speed
\begin{equation}\label{def_beta}
	\beta \equiv \frac{\alpha}{\gamma}.
\end{equation}
It follows that
\begin{equation}
	\beta=\sqrt{1-f},
	\qquad
	\alpha=\gamma\beta,
	\qquad
	\gamma=\frac{1}{\sqrt{1-\beta^2}}.
\end{equation}
Thus, \(\beta\) is the local speed of the free-falling frame with respect to
the static frame. Since the boost direction is given by the radial unit
vector in Eq.~\eqref{radial_unit_vector}, the corresponding local velocity
may be written as
\begin{equation}
	\boldsymbol{\beta}=\beta\,\mathbf{n}.
\end{equation}
The sign of the radial velocity depends on the orientation convention for
the local radial basis. For the geodesic used here, the motion is inward in
Schwarzschild coordinates, as indicated by the radial component of
Eq.~\eqref{u_freefalling}. The local Lorentz transformation itself,
however, is completely fixed by the tetrads through
Eq.~\eqref{lorentz_transformation}.

For the Schwarzschild function \(f(r)=1-2M/r\), the relative speed becomes
\begin{equation}
	\beta(r)=\sqrt{\frac{2M}{r}}.
\end{equation}
This is the escape velocity, or equivalently the velocity acquired by a
particle falling radially from rest at infinity, as measured by a local
static observer. Therefore, the free-falling frame can be viewed as the
local inertial frame obtained from the static one by a radial boost with
speed \(\sqrt{2M/r}\).

This point is central for the interpretation of the Maxwell equations in
Sec.~\ref{sec5}. The additional terms appearing in the free-falling
equations are not new gravitational sources. Rather, they arise because the
free-falling observer decomposes the same tensorial equations with a
different local time direction, i.e., the transformation from the
static frame to the free-falling frame changes the local split
\begin{equation}
	F_{\mu\nu}
	\longrightarrow
	(\mathbf{E},\mathbf{B}),
	\qquad
	J^\mu
	\longrightarrow
	(\rho,\mathbf{J}).
\end{equation}
Consequently, quantities which are purely electric, magnetic, temporal or
spatial in one frame need not remain so in the other.

The four-current provides the most transparent illustration. As shown in
Eq.~\eqref{current_boost_spherical}, the boost mixes only the temporal and
radial components, whereas the angular current components remain unchanged.
Equivalently, one may write the temporal-radial sector as
\begin{equation}
		\begin{pmatrix}
			\bar \rho\\[0.3em]
			\bar{J}_{r}
	\end{pmatrix}
	=
	\begin{pmatrix}
		\gamma & \alpha\\
		\alpha & \gamma
	\end{pmatrix}
	\begin{pmatrix}
		\tilde{\rho}\\[0.3em]
		\tilde{J}_{r}
	\end{pmatrix}.
\end{equation}
Therefore, charge density and radial current are observer-dependent parts
of the same four-vector. A source at rest in the static frame,
\(\tilde{J}_{r}=0\), is measured by the free-falling observer as a source with
both charge density and radial current. From Eq.~\eqref{current_boost_spherical},
one obtains
\begin{equation}
	\frac{\bar{J}_{r}}{\bar{\rho}}
	=
	\frac{\alpha}{\gamma}
	=
	\beta
	=
	\sqrt{1-f}.
\end{equation}
Thus, the free-falling observer interprets a static charged distribution as
a distribution moving radially with the relative speed \(\beta\). This is
the usual special-relativistic mixing between charge density and current,
now applied locally in Schwarzschild spacetime.

Conversely, a radial current with vanishing charge density in the static
frame gives a non-vanishing charge density in the free-falling frame, again
as follows directly from Eq.~\eqref{current_boost_spherical}. This does not
represent charge creation. It is only the observer dependence of the
temporal component of the four-current.

A similar interpretation applies to the electromagnetic fields. The fields
measured by an observer with four-velocity \(u^a\) are obtained by
projecting the electromagnetic tensor and its dual according to
\begin{equation}
	E^a=F^a{}_{b}u^b,
	\qquad
	B^a=\mathcal{F}^a{}_{b}u^b .
\end{equation}
Changing from the static four-velocity to the free-falling one changes the
projection and therefore changes the measured electric and magnetic fields.
For a local boost along the radial direction, the components parallel to
the boost are distinguished from the transverse ones. Schematically,
\begin{equation}
	E_\parallel \ \hbox{and}\ B_\parallel
	\quad \hbox{are not mixed,}
\end{equation}
whereas the transverse field components may mix as
\begin{equation}
	\mathbf{E}_{\perp}^{\,\rm ff}
	\sim
	\gamma
	\left(
		\mathbf{E}_{\perp}^{\,\rm stat}
		+
		\boldsymbol{\beta}\times
		\mathbf{B}_{\perp}^{\,\rm stat}
	\right),
\end{equation}
and
\begin{equation}
	\mathbf{B}_{\perp}^{\,\rm ff}
	\sim
	\gamma
	\left(
		\mathbf{B}_{\perp}^{\,\rm stat}
		-
		\boldsymbol{\beta}\times
		\mathbf{E}_{\perp}^{\,\rm stat}
	\right),
\end{equation}
up to the sign convention fixed by the orientation of the radial boost.
This is the local special-relativistic mixing of electric and magnetic
fields. In the present problem, however, the boost parameter depends on the
radial position through \(f(r)\).
This statement concerns the field components measured by the two observers;
it does not by itself imply that every projected Maxwell equation changes
form. In fact, as shown explicitly in Sec.~\ref{sec5}, the angular
Amp\`ere-Maxwell and Faraday equations remain the same as in the static
frame.

The radial dependence of the boost is nevertheless important for the
temporal-radial projections of the equations. It should not, however, be
confused with the geometrical terms already present in the static frame.
In the angular components of Amp\`ere-Maxwell law and Faraday's law, the
free-falling equations contain the same combinations
\(\partial_r(rf^{1/2}X_A)\), with \(A=\theta,\phi\), that appear for the
static observer. If these derivatives are expanded, the resulting terms
proportional to \(f'\) are the same geometrical lapse-variation terms
discussed in the static case, not additional contributions produced by the
free-fall boost.

The free-falling equations therefore contain two distinct types of effects.
The first is the same geometrical effect already present in the static
frame, associated with the Schwarzschild radial and temporal metric
factors. The second is the kinematical effect of a position-dependent radial
boost. It is this second effect that produces the mixing between charge and
radial current, and between the electric and magnetic sectors in the
temporal-radial projections of Maxwell's equations. The angular
Amp\`ere-Maxwell and Faraday equations, by contrast, remain identical in
form to the corresponding equations measured by the static observer. This
distinction will be important in the following subsection, where the mixed
structure of the free-falling equations is analyzed in detail.

\subsection{Mixing of electric, magnetic, charge and current sectors}

The previous subsection shows that the free-falling frame is locally related
to the static frame by a radial Lorentz boost. We now analyze the main
physical consequence of this fact. Since the electric and magnetic fields
are not independent tensorial objects, but observer-dependent projections of
the electromagnetic tensor, the passage from the static frame to the
free-falling frame changes the split between electric and magnetic sectors.
In the Maxwell equations obtained above, this change appears in the
temporal-radial projections. The same happens to the four-current, whose
temporal and radial components are mixed by the radial boost.

Let us begin with the sources. According to
Eq.~\eqref{current_boost_spherical}, the charge density and the radial
current measured by the free-falling observer are given by a Lorentz mixing
of the corresponding quantities measured in the static frame. Therefore,
neither the charge density nor the radial current has an invariant meaning
by itself. They are the temporal and radial projections of the same
four-current.

This fact has an immediate physical interpretation. As already follows from
Eq.~\eqref{current_boost_spherical}, a source at rest with respect to the
static observer is measured by the free-falling observer as a source with a
radial convective current. Conversely, a radial current with vanishing charge
density in the static frame may have a non-vanishing charge density in the
free-falling frame. Both express the observer dependence of the temporal and radial projections of \(J^\mu\).

The same mechanism explains the structure of the inhomogeneous Maxwell
equations in the free-falling frame. In the static frame, Gauss's law,
Eq.~\eqref{Gauss_static}, contains only the electric divergence and the
charge density. By contrast, the free-falling Gauss's law,
Eq.~\eqref{Gauss_freefalling}, contains electric terms, magnetic angular
terms and the time derivative of the radial electric field. This is not a
new electromagnetic coupling. It is the boost mixing of the temporal
projection of Maxwell's equations with its radial projection.

This statement can be made more explicit by arranging the static equations
in source-isolated form. Let
\begin{equation}
	{\cal G}_{E}^{\rm stat}=\tilde{\rho}
\end{equation}
denote the static Gauss's law, Eq.~\eqref{Gauss_static}, and let
\begin{equation}
	{\cal A}_{r}^{\rm stat}=\tilde{J}_{r}
\end{equation}
denote the radial component of Amp\`ere-Maxwell law, Eq.~\eqref{AMr},
with the displacement-current term placed on the left-hand side. Then the
temporal-radial sector of the inhomogeneous equations in the free-falling
frame has the same boost structure as the four-current, i.e.,
\begin{equation}
	\begin{pmatrix}
		{\cal G}_{E}^{\rm ff}\\[0.3em]
		{\cal A}_{r}^{\rm ff}
	\end{pmatrix}
	=
	\begin{pmatrix}
		\gamma & \alpha\\
		\alpha & \gamma
	\end{pmatrix}
	\begin{pmatrix}
		{\cal G}_{E}^{\rm stat}\\[0.3em]
		{\cal A}_{r}^{\rm stat}
	\end{pmatrix},
\end{equation}
where \({\cal G}_{E}^{\rm ff}\) and \({\cal A}_{r}^{\rm ff}\) are the
left-hand sides of Eqs.~\eqref{Gauss_freefalling} and
\eqref{AMr_freefalling}, respectively. Consequently,
\begin{equation}
	{\cal G}_{E}^{\rm ff}
	=
	\gamma\tilde{\rho}+\alpha\tilde{J}_{r},
	\qquad
	{\cal A}_{r}^{\rm ff}
	=
	\alpha\tilde{\rho}+\gamma\tilde{J}_{r},
\end{equation}
in agreement with the source terms in
Eqs.~\eqref{Gauss_freefalling} and \eqref{AMr_freefalling}.

This form is physically important. It shows that what appears as Gauss's law
for the free-falling observer is not simply Gauss's law in the static
observer's frame written in different coordinates. Rather, it is a Lorentz mixture
of the static Gauss's law and the radial Amp\`ere-Maxwell law. Thus, a
constraint equation for one observer may contain part of an evolution
equation for another observer. This is not a violation of the constraint
structure of Maxwell theory, but a consequence of changing the local time
direction used to decompose the covariant equations.

The angular components have a different behavior. Since the boost is
radial, the angular components of the current are not mixed:
\begin{equation}
	\bar{J}_{\theta}=\tilde{J}_{\theta},
	\qquad
	\bar{J}_{\phi}=\tilde{J}_{\phi} .
\end{equation}
Accordingly, the source terms in
Eqs.~\eqref{AMtheta_freefalling} and \eqref{AMphi_freefalling} remain
\(\tilde{J}_{\theta}\) and \(\tilde{J}_{\phi}\), respectively. More than
that, the angular Amp\`ere-Maxwell equations themselves coincide with
Eqs.~\eqref{AMtheta} and \eqref{AMphi}, once both are written with the
time-derivative terms on the same side. Thus the polar and azimuthal
inhomogeneous equations are not part of the boost-induced mixing sector.
Any terms proportional to \(f'\) that appear after expanding the radial
derivatives in these angular equations have the same geometrical origin as
in the static frame.

The homogeneous equations exhibit the same observer-dependent separation.
In the static frame, the magnetic Gauss's law, Eq.~\eqref{GaussB_static},
contains only the divergence of the magnetic field. In the free-falling
frame, Eq.~\eqref{GaussB_freefalling} contains magnetic terms, electric
angular terms and the time derivative of the radial magnetic field. This
does not indicate the presence of magnetic charges. The equation still
expresses the absence of magnetic monopoles, but in the electromagnetic
decomposition performed by the free-falling observer.

In analogy with the inhomogeneous sector, let
\begin{equation}
	{\cal G}_{B}^{\rm stat}=0
\end{equation}
denote the static magnetic Gauss's law, Eq.~\eqref{GaussB_static}, and let
\begin{equation}
	{\cal F}_{r}^{\rm stat}=0
\end{equation}
denote the radial component of Faraday's law, Eq.~\eqref{Faradayr}, written
in the conventional form with the time-derivative term placed on the left-hand
side. It is useful, however, to distinguish this conventional Faraday equation
from the radial component of the projected homogeneous equation. If
\begin{equation}
	{\cal H}^{b}\equiv
	\partial_{\mu}\left(e\,{\cal F}^{\mu b}\right)
	+
	e\,{\cal F}^{\mu c}\mathring{\omega}_{\mu}{}^{b}{}_{c}
\end{equation}
denotes the left-hand side of the projected homogeneous equations, then, with
the conventions adopted in this work,
\begin{equation}
	{\cal H}^{(0)}={\cal G}_{B},
	\qquad
	{\cal H}^{(r)}=-{\cal F}_{r}.
\end{equation}
Thus the temporal-radial components of the homogeneous equations transform
with the usual boost matrix,
\begin{equation}
	\begin{pmatrix}
		{\cal H}_{\rm ff}^{(0)}\\[0.3em]
		{\cal H}_{\rm ff}^{(r)}
	\end{pmatrix}
	=
	\begin{pmatrix}
		\gamma & \alpha\\
		\alpha & \gamma
	\end{pmatrix}
	\begin{pmatrix}
		{\cal H}_{\rm stat}^{(0)}\\[0.3em]
		{\cal H}_{\rm stat}^{(r)}
	\end{pmatrix}.
\end{equation}
The radial component of Faraday's law therefore
participates in the same temporal-radial mixing, while the angular components
of Faraday's law remain identical in form to Eqs.~\eqref{Faradaytheta} and
\eqref{Faradayphi}.

The origin of all these effects is the observer-dependent definition of the
electric and magnetic fields discussed above. Changing the observer changes
the local split of \(F_{ab}\) into electric and magnetic parts, while the
tensor \(F_{ab}\) itself is unchanged. Consequently, magnetic contributions
in the free-falling Gauss's law do not mean that magnetism sources electric
charge; electric contributions in the magnetic Gauss's law do not mean that
electric fields source magnetic charge. They only show that the
electric-magnetic split has changed.

The invariant content of the electromagnetic field remains encoded in
observer-independent scalars, such as
\begin{equation}
	I_1=F_{ab}F^{ab},
	\qquad
	I_2=F_{ab}\mathcal{F}^{ab}.
\end{equation}
In terms of locally measured fields, these invariants correspond, up to
conventional signs, to
\begin{equation}
	I_1\sim B^2-E^2,
	\qquad
	I_2\sim \mathbf{E}\cdot\mathbf{B}.
\end{equation}
Therefore, although the measured electric and magnetic fields depend on the
observer, the tensorial electromagnetic field does not.

The results of Secs.~\ref{sec4} and \ref{sec5} should therefore be read as
two different local decompositions of the same covariant Maxwell equations.
The static observer sees geometrical corrections in the radial and temporal
operators, but no electric-magnetic mixing. The free-falling observer sees,
in addition, the kinematical mixing induced by the radial boost in the
temporal-radial sector. The angular Amp\`ere-Maxwell and Faraday equations
remain the same as for the static observer. In this sense, the free-falling
equations make explicit that the division into
electric field, magnetic field, charge density and current density is not an
absolute property of the electromagnetic configuration, but a property of
the observer used to measure it.

\subsection{Weak-field and near-horizon regimes}

The interpretation developed in the previous subsections becomes especially
transparent in two limiting regimes: the weak-field region, \(r\gg M\),
and the near-horizon region, \(r\to 2M\). These limits clarify the
relative role of the geometrical corrections already present in the static
frame and of the kinematical mixing introduced by the radially
free-falling frame.

Let us first consider the weak-field regime. For
\begin{equation}
	f(r)=1-\frac{2M}{r},
\end{equation}
with \(r\gg M\), one has
\begin{equation}
	f^{1/2}
	=
	1-\frac{M}{r}
	+O\left(\frac{M^2}{r^2}\right),
\end{equation}
and
\begin{equation}
	f^{-1/2}
	=
	1+\frac{M}{r}
	+O\left(\frac{M^2}{r^2}\right).
\end{equation}
Therefore, the geometrical corrections appearing in the static equations
are of order \(M/r\). This is the expected order of the gravitational
potential in the weak-field approximation. In this sense, the static
observer sees Maxwell equations whose radial and temporal differential
operators are deformed perturbatively by the Schwarzschild geometry.

The situation is different for the free-falling frame. From the boost
parameters introduced in Eq.~\eqref{boost_parameters}, and from the
definition \(\beta=\alpha/\gamma\) given by \eqref{def_beta}.
Thus, the free-fall velocity measured by the local static observer is of
order
\begin{equation}\label{beta_schwarzschild}
	\beta\sim \left(\frac{M}{r}\right)^{1/2}.
\end{equation}
Consequently, the mixing effects associated with the radial boost enter at
kinematical order \(\beta\), whereas the purely geometrical corrections of
the static frame enter at gravitational order \(M/r\).

This distinction is physically relevant. In the weak-field region, the
static-frame corrections are small because the metric differs only slightly
from the Minkowski metric. However, the free-falling observer may still
measure electric-magnetic and charge-current mixing terms of order
\(\sqrt{M/r}\) in the temporal-radial projections. Hence, whenever radial
currents are present, or whenever the temporal-radial equations involve
angular curls or divergences of transverse fields, the leading difference
between the static and free-falling measurements can be governed by the
relative velocity between the observers, rather than directly by the metric
correction itself.

Equivalently, the weak-field limit separates two effects:
\begin{equation}
\begin{aligned}
	\hbox{geometrical corrections}
	&\sim
	O\left(\frac{M}{r}\right),
	\\
	\hbox{boost mixing}
	&\sim
	O\left(\sqrt{\frac{M}{r}}\right).
\end{aligned}
\end{equation}
In this regime, the free-falling equations
are therefore close to the special-relativistic transformation of Maxwell
equations under a small radial boost, supplemented by weak gravitational
corrections.

The near-horizon regime has a different character. As
\begin{equation}
	r\to 2M,
	\qquad
	f(r)\to 0,
\end{equation}
the boost parameters behave as
\begin{equation}
	\gamma=f^{-1/2}\to \infty,
	\qquad
	\alpha=\sqrt{\frac{1}{f}-1}\to \infty,
\end{equation}
whereas
\begin{equation}
	\beta=\sqrt{1-f}\to 1 .
\end{equation}
Thus, the free-falling observer approaches the speed of light with respect
to the local static frame. This statement must be interpreted with care:
the divergence of \(\gamma\) and \(\alpha\) does not signal a physical
singularity of the free-falling observer. Instead, it signals the breakdown
of the static frame at the Schwarzschild radius.

A static observer requires a non-vanishing proper acceleration in
order to remain at fixed \(r\). As discussed above, this proper acceleration
is proportional to
\begin{equation}
	a_{\rm stat}
	\sim
	\frac{1}{f^{1/2}},
\end{equation}
and therefore diverges as \(f\to 0\). Hence, the static frame becomes
singular at the horizon. The divergent boost between the free-falling and
static frames is a manifestation of this fact. It is not, by itself, an
indication that the electromagnetic field measured by a regular
free-falling observer diverges.

This distinction is important for the interpretation of the free-falling
Maxwell equations. Since the free-falling equations were written in terms of
fields and sources explicitly related to the static decomposition, factors
such as \(\gamma\), \(\alpha\), \(f^{-1/2}\), and \(f^{-1}\) may appear in
the expressions. Near the horizon, these factors reflect the increasingly
singular relation between the static and free-falling frames. A regularity
analysis of the electromagnetic field should therefore be formulated in
terms of quantities measured directly by a regular observer, or in terms of
tensorial invariants.

The weak-field and near-horizon limits thus emphasize complementary aspects
of the same result. Far from the black hole, the static equations differ
from their flat-spacetime form by small geometrical corrections, while the
free-falling equations already display kinematical mixing at order
\(\sqrt{M/r}\) in the temporal-radial sector. Near the horizon, the relative
boost between static and free-falling observers becomes singular because
the static frame ceases to be physically admissible.

\subsection{The gravitational field as an effective medium}

For the static observers, the Schwarzschild gravitational field behaves as
an inhomogeneous effective medium at rest. In this frame, the equations
obtained in Sec.~\ref{sec4} contain metric corrections in the radial and
temporal sectors, but they do not mix electric and magnetic fields. Thus,
the static observer sees a geometrical modification of the vacuum Maxwell
equations, rather than a magnetoelectric coupling.

This is already visible in Gauss's law. The radial part of
Eq.~\eqref{Gauss_static} contains the factor \(f^{1/2}\), while the angular
sector keeps the usual spherical form. One may therefore say that the
gravitational field modifies the radial electric response of the vacuum, in
a manner analogous to an inhomogeneous medium. However, this interpretation
is not unique. The metric factors may be regarded either as part of the
differential operators, as shown previously, or as part of effective constitutive relations.

This distinction is important. A naive identification such as
\(D_r=f^{1/2}\tilde E_r\) would not reproduce directly the radial part of
Eq.~\eqref{Gauss_static}, because
\begin{equation}
	\frac{1}{r^2}\partial_r\left(r^2 f^{1/2}\tilde E_r\right)
	\neq
	\frac{f^{1/2}}{r^2}\partial_r\left(r^2\tilde E_r\right).
\end{equation}
The two expressions differ by a term proportional to \(f'\tilde E_r\).
Thus, when written in terms of locally measured tetrad components, the
effective-medium interpretation should be used with care: the geometry can
be absorbed into effective material coefficients only after a precise
choice of electromagnetic variables.

The same observation applies to the magnetic sector. In
Eqs.~\eqref{AMtheta}--\eqref{AMphi} and
Eqs.~\eqref{Faradaytheta}--\eqref{Faradayphi}, the radial derivatives of
the transverse fields appear through combinations involving
\(r f^{1/2}\tilde B_A\) and \(r f^{1/2}\tilde E_A\), with
\(A=\theta,\phi\). Therefore, the gravitational field affects both electric
and magnetic sectors. The analogy is not merely with a dielectric medium,
but with a geometrical medium whose electric and magnetic responses are
both controlled by the same metric function \(f(r)\).

For the free-falling observers, the interpretation changes in the
temporal-radial sector. Since the free-falling tetrad is related to the
static tetrad by a radial boost, the same effective medium is seen in radial
motion in that sector. In ordinary electrodynamics, a moving material medium
produces magnetoelectric-like couplings: electric and magnetic responses are
mixed by the relative motion. Here, an analogous effect occurs in the
temporal-radial projections. The free-falling equations contain magnetic
terms in the electric Gauss's law, electric terms in the magnetic Gauss's
law, and a mixing between charge density and radial current. The angular
Amp\`ere-Maxwell and Faraday equations, however, retain the same form as in
the static frame.

Hence, the analogy may be summarized as follows: the static frame
corresponds to an effective medium at rest, whereas the free-falling frame
corresponds to a radially moving effective medium in the temporal-radial
sector.
The first picture accounts for the geometrical factors in the static
equations. The second accounts for the boost-induced mixing in the
free-falling temporal-radial equations.

In order to illustrate the static medium analogy, let us first consider a purely radial electric configuration,
\begin{equation}
	\tilde{\mathbf E}
	=
	\tilde E_r(r)\,\hat{\mathbf r},
	\qquad
	\tilde{\mathbf B}=0 .
\end{equation}
This example is useful because the electric field is longitudinal with
respect to the boost relating the static and free-falling frames.

For the static observer, Eq.~\eqref{Gauss_static} reduces to
\begin{equation}
	\frac{f^{1/2}}{r^2}
	\frac{d}{dr}\left(r^2\tilde E_r\right)
	=
	\tilde{\rho} .
\end{equation}
In vacuum, this gives
\begin{equation}
	\frac{d}{dr}\left(r^2\tilde E_r\right)=0,
\end{equation}
and therefore
\begin{equation}
	\tilde E_r=\frac{Q}{r^2},
\end{equation}
where \(Q\) is an integration constant. Thus, outside the source, the
locally measured radial electric field has the usual inverse-square
dependence.

The effective-medium interpretation appears when a source is present. If
one rewrites the previous equation using the ordinary spherical radial
divergence, one obtains
\begin{equation}
	\frac{1}{r^2}
	\frac{d}{dr}\left(r^2\tilde E_r\right)
	=
	\frac{\tilde{\rho}}{f^{1/2}} .
\end{equation}
In this form, the gravitational field effectively rescales the source that
appears in the flat-space-like radial divergence. This resembles the role
played by a material medium in Gauss's law. 

The same configuration also provides a consistency check for the
free-falling equations. If the field is static and purely radial, with
\(\tilde{\mathbf B}=0\), the magnetic terms in Eq.~\eqref{Gauss_freefalling}
vanish. For a source at rest in the static frame, \(\tilde{J}_{r}=0\), the
right-hand side of Eq.~\eqref{Gauss_freefalling} becomes
\(\gamma\tilde{\rho}\). Using the static Gauss's law, the left-hand side becomes
\begin{equation}
	\frac{1}{r^2}
	\frac{d}{dr}\left(r^2\tilde E_r\right)
	=
	\frac{\tilde{\rho}}{f^{1/2}}
	=
	\gamma\tilde{\rho} .
\end{equation}
Thus, the free-falling Gauss's law is satisfied without the appearance of an
additional magnetic sector. This confirms that longitudinal fields are not
magnetoelectrically mixed by a longitudinal boost.

\section{Conclusions}\label{conc}
In this work, we investigated Maxwell equations in Schwarzschild spacetime
in the perspective of different families of observers. The main idea
was to use the tetrad formulation in which the electromagnetic fields and
sources are projected onto the local frame of a given observer. In this
approach, the electromagnetic tensor and the four-current remain the
fundamental covariant objects, while the quantities interpreted as electric
field, magnetic field, charge density and current density are local
measurements performed by the chosen frame.

The analysis was based on the formulation of Maxwell equations in arbitrary
frames developed by Maluf and Ulhoa. This formalism is especially useful because it
allows one to change the physical reference frame without changing the
spacetime coordinate system. Hence, the effects associated with the observer
can be computed directly at the tetrad level, avoiding mixing genuine
frame effects with coordinate artifacts. In the present case, both the
static and the radially free-falling observers were described in the same
Schwarzschild coordinates, and the differences between their measured
Maxwell equations were entirely encoded in the frame of reference.

For the static observer, Maxwell equations preserve the usual spherical
structure. The gravitational field appears through the factors \(f^{1/2}\)
and \(f^{-1/2}\), which modify the radial and temporal sectors of the
equations, respectively. These factors may be interpreted geometrically in
terms of proper radial distance and proper time. Thus, in the static frame,
Schwarzschild geometry acts mainly as a geometrical deformation of the
flat-spacetime spherical Maxwell equations.

For the radially free-falling observer, the situation is richer. The
free-falling frame is related to the static one by a local radial boost, as
shown by Eqs.~\eqref{Lambda_radial_boost}--\eqref{boost_parameters}. As a
result, the free-falling equations contain both the geometrical corrections
already present in the static frame and the kinematical effects associated
with this boost. In particular, charge density mixes with radial current,
as shown in Eq.~\eqref{current_boost_spherical}, and the electric and
magnetic sectors are also mixed in the temporal-radial projections of
Maxwell equations. The angular components of Amp\`ere-Maxwell and Faraday's
laws, on the other hand, retain the same form found for the static observer.

This mixing does not represent the appearance of new electromagnetic
sources. The magnetic terms in the free-falling electric Gauss's law,
Eq.~\eqref{Gauss_freefalling}, and the electric terms in the free-falling
magnetic Gauss's law, Eq.~\eqref{GaussB_freefalling}, arise because different
observers perform different local decompositions of the same tensor
\(F_{\mu\nu}\). Similarly, the mixing between charge density and radial
current reflects the observer dependence of the decomposition of the same
four-current \(J^\mu\).

The weak-field and near-horizon analyses reinforce this interpretation. In
the weak-field regime, the geometrical corrections associated with the
static frame are of gravitational order, whereas the free-fall mixing in the
temporal-radial sector is of kinematical order, controlled by the relative
velocity between the observers. Near the horizon, the boost relating the
static and free-falling frames becomes singular, reflecting the breakdown of
the static frame at the Schwarzschild radius, not a physical singularity of
the free-falling description.

We also discussed an effective-medium interpretation of the results. For
static observers, the Schwarzschild gravitational field behaves
operationally as an inhomogeneous geometrical medium at rest, while the
free-falling observer sees the corresponding radially moving-medium behavior
in the temporal-radial sector.

The results illustrate the power of the arbitrary-frame formulation. Once
the equations are written in tetrad components, one may change the physical
observer without changing coordinates. The same strategy can be applied to
rotating observers, accelerated frames, locally non-rotating observers in
axisymmetric spacetimes, or any other tetrad adapted to a chosen family of
observers. In each case, the formalism provides a direct way to identify
which terms arise from spacetime geometry and which terms arise from the
state of motion of the observer.



\begin{thebibliography}{10}

\bibitem{minkowski1908grundgleichungen}
H.~Minkowski.
\newblock Die grundgleichungen f{\"u}r die elektromagnetischen vorg{\"a}nge in
  bewegten k{\"o}rpern.
\newblock {\em Nachrichten von der Gesellschaft der Wissenschaften zu
  G{\"o}ttingen, Mathematisch-Physikalische Klasse}, 1908:53--111, 1908.

\bibitem{hehl2008maxwell}
F.~W. Hehl.
\newblock Maxwell's equations in minkowski's world: their premetric
  generalization and the electromagnetic energy-momentum tensor.
\newblock {\em Annalen der Physik}, 520(9-10):691--704, 2008.

\bibitem{hwang2023definition}
J.~C. Hwang and H.~Noh.
\newblock Definition of electric and magnetic fields in curved spacetime.
\newblock {\em Annals of Physics}, 454:169332, 2023.

\bibitem{aly2025coupling}
F.~Aly and D.~Stojkovic.
\newblock Coupling between gravitational and electromagnetic perturbations on
  kerr spacetime.
\newblock {\em arXiv preprint arXiv:2511.13642}, 2025.

\bibitem{frolov2020maxwell}
V.~P. Frolov.
\newblock Maxwell equations in a curved spacetime: Spin optics approximation.
\newblock {\em Physical Review D}, 102(8):084013, 2020.

\bibitem{mavrogiannis2021electromagnetic}
P.~Mavrogiannis and Christos~G. Tsagas.
\newblock Electromagnetic potentials in curved spacetimes.
\newblock {\em Classical and Quantum Gravity}, 38(23):235002, 2021.

\bibitem{formiga2023gravitational}
J.~B. Formiga and J.~A.~C. Duarte.
\newblock Gravitational energy problem and the energy of photons.
\newblock {\em Physical Review D}, 108(4):044043, 2023.

\bibitem{weitzenbock1932invarianten}
R.~Weitzenb{\"o}ck.
\newblock {\"U}ber die invarianten von linearen gruppen, 1932.

\bibitem{maluf2013teleparallel}
J.~W. Maluf.
\newblock The teleparallel equivalent of general relativity.
\newblock {\em Annalen der Physik}, 525(5):339--357, 2013.

\bibitem{maluf2010electrodynamics}
J.~W. Maluf and S.~C. Ulhoa.
\newblock Electrodynamics in accelerated frames revisited.
\newblock {\em Annalen der Physik}, 522(10):766--775, 2010.

\bibitem{bremm2015nonlocal}
G.~N. Bremm and F.~T. Falciano.
\newblock Nonlocal effects in black body radiation.
\newblock {\em Annalen der Physik}, 527(3-4):265--277, 2015.

\bibitem{obukhov2021electrodynamics}
Y.~N. Obukhov.
\newblock Electrodynamics in noninertial frames.
\newblock {\em The European Physical Journal C}, 81(10):919, 2021.

\bibitem{formiga2014accelerated}
J.~B. Formiga.
\newblock On the accelerated observers proper coordinates and the rigid motion
  problem in minkowski spacetime.
\newblock {\em Brazilian Journal of Physics}, 44(1):95--101, 2014.

\bibitem{maluf2005gravitational}
J.~W. Maluf.
\newblock The gravitational energy-momentum tensor and the gravitational
  pressure.
\newblock {\em Annalen der Physik}, 517(11-12):723--732, 2005.

\bibitem{maluf2020difficulties}
J.~W. Maluf, S.~C. Ulhoa, J.~F. da~Rocha-Neto, and F.~L. Carneiro.
\newblock Difficulties of teleparallel theories of gravity with local lorentz
  symmetry.
\newblock {\em Classical and Quantum Gravity}, 37(6):067003, 2020.

\bibitem{mashhoon2003vacuum}
B.~Mashhoon.
\newblock Vacuum electrodynamics of accelerated systems: Nonlocal maxwell's
  equations.
\newblock {\em Annalen der Physik}, 515(10):586--598, 2003.

\end{thebibliography}
\end{document}